\documentclass[final,5p,times,twocolumn]{elsarticle}

\usepackage{amsmath,amssymb,amsfonts}
\usepackage{algorithmic}
\usepackage{algorithm}
\usepackage{array}
\usepackage{graphicx}
\usepackage{textcomp}
\usepackage{stfloats}
\usepackage{xcolor}
\usepackage{float}
\usepackage{tcolorbox}
\usepackage{enumitem}
\usepackage{hyperref}
\usepackage{multirow}
\usepackage{url}
\usepackage{graphicx}
\usepackage{subfigure} 
\usepackage{booktabs}
\usepackage{textcomp}
\usepackage{caption}
\usepackage{balance}

\usepackage{lineno}
\linenumbers

\newcommand{\yj}[1]{\textcolor{black}{#1}}
\newcommand{\yjnew}[1]{\textcolor{black}{#1}}

\journal{Journal of Systems and Software}

\begin{document}
\nolinenumbers
\begin{frontmatter}



\title{An Exploratory Study on LLM-Generated Code and Comments in Code Repositories}


\author{Yongyi Ji}
\author{Jiaji Wang}
\author{Yi Zhou}
\author{Fuxiang Chen \corref{cor1}}
\author{Hongji Yang} 
\cortext[cor1]{Corresponding author: Fuxiang Chen (email: fuxiang.chen@leicester.ac.uk)}
\affiliation{organization={School of Computing and Mathematical Sciences, University of Leicester},
            addressline={University Road}, 
            city={Leicester},
            postcode={LE1 7RH}, 
            country={United Kingdom}}

\begin{abstract}
The use of LLMs in software development has become increasingly widespread on tasks such as code generation and summarization. Reports from large technology companies showed that around 20\% to 30\% of their code are generated by LLMs. However, there remains skepticism about the practical usage of LLM-generated code and comments, such as concerns on more time for debugging the generated code and the unnaturalness of the generated comments. 
In this paper, we study the code and comments \yjnew{detected as }likely to be generated by LLMs and their characteristics, the differences between company- and community-maintained repositories, and how likely bugs are \yj{associated with LLM-generated code.}
We conduct extensive experiments on active company- and community-maintained repositories from 2021 to 2025 using various tools and techniques that detect code and comments generated by LLMs. 
\yjnew{Based on our detector-based proxy analysis, the results suggest that} code detected as likely to be generated by LLMs decreased over time and \yjnew{appeared frequently in} test cases, while that of comments remains relatively \yjnew{stable}. \yjnew{Proxy results further suggest that code detected as likely to be generated by LLMs shows substantial intra-repository code clones, whereas comments exhibit a relatively low proportion of grammatically correct sentences. In addition}, the company-maintained repositories show a higher percentage of code and comments detected as likely to be generated by LLMs, and only a small percentage of the human-labelled bugs 
are detected as being likely \yj{associated with LLM-generated code}. 

\end{abstract}


\end{frontmatter}



\section{Introduction}
\label{section:introduction}



The rapid advancement of Large Language Models (LLMs) has resulted in multiple adaptation of LLMs in various downstream tasks \cite{zhao2023survey}. In Software Engineering (SE), it was reported that some developers are already using LLMs in code generation and summarization, the two most important activities in software development and comprehension \cite{weisz2025examining,ciniselli2021empirical, ahmed2022few, jiang2024survey, shojaee2023execution, sun2024source, lu2021codexglue}. Moreover, the JetBrains Developer Ecosystem Survey \cite{JetBrains} found that developers use LLMs to generate code and comments frequently. There have also been reports suggesting that the use of LLMs is gaining popularity among software companies. For example, in 2024, Google stated that 25\% of the company’s new code is generated by LLMs \cite{google-ceo}. Similarly, in 2025, Microsoft stated that 20\% to 30\% of the code within the company’s repositories is written by LLMs \cite{microsoft-ceo}. In addition, the 2025 Google’s DORA report \cite{dora-report}, which surveyed nearly 5,000 professional developers from around the world, found that 90\% of the surveyed developers reportedly use some elements of LLMs at work, and over 80\% of them believed this had improved their productivity. 



On the other hand, there is reported skepticism on the accuracy of LLM-generated content. For code, Pearce et al. \cite{10.1145/3610721} found that when Github Copilot \cite{copilot-github} is prompted to generate code in security-related scenarios, 40\% of the code generated by Github Copilot was vulnerable. In the paper, the Github Copilot refers to the Copilot AI service hosted by Github. This is not to be mistaken for the Copilot model developed by Microsoft \cite{copilot-microsoft}. Liang et al. \cite{liang2024large} surveyed 410 developers and found that the primary reason for not using LLMs was that the generated code did not meet their requirements. 
For comments, Sergeyuk et al. \cite{sergeyuk2025using} collected opinions from 481 developers and found that one of the top reasons for not using comments generated by LLMs is that the generated comments are often unnatural and that they do not match the required tone and clarity of a human developer. Moreover, the code and comments that were generated by LLMs raised the uncertainty on whether a generated software artifact is created by an individual or a machine \cite{10.1109/ICSE55347.2025.00005, 10.1145/3605770.3625215}.

In light of the controversial reports on the use of LLMs by developers, there are multiple unknowns: 1) What proportion of the code and comments in developers' repositories are likely to be generated by LLMs? 2) What characteristics do these code and comments possess? 3) Are there any differences between company- and community-maintained repositories on these code and comments? 4) Are there bugs in repositories that were \yj{associated with LLM-generated code?}
By answering these questions, we can better establish the gaps between developers’ LLMs usages and the
current state of LLMs. \yj{Previous work have proposed different detectors to detect text that is likely generated by LLMs \cite{gehrmann2019gltr, mitchell2023detectgpt, bao2023fast}. Our work differs from previous research by applying existing detectors to repositories in order to quantify the proportion of code and comments that are likely to be generated by LLMs and to analyze their characteristics, rather than proposing a new detector. }
In this study, we attempt to answer these questions by focusing on detecting code and comments that are likely to be generated by LLMs in the developers' repositories as writing code and comments are the two most important activities in software development as mentioned earlier. From this point onwards, we use likely to be LLM-generated to refer to code or comments that are likely to be generated by LLMs. \yjnew{We emphasize that all our findings are proxy-based observations derived from detector outputs, given the absence of ground-truth labels.}

In our experiments, we apply existing LLM-generated content detectors, including Binoculars \cite{10.5555/3692070.3692768}, Log-Likelihood \cite{solaiman2019release}, Entropy \cite{lavergne2008detecting}, Rank \cite{gehrmann2019gltr}, Log-Rank \cite{gehrmann2019gltr}, LRR \cite{su-etal-2023-detectllm}, 
DetectGPT \cite{mitchell2023detectgpt}, Fast-DetectGPT \cite{bao2023fast} and DetectCodeGPT \cite{10.1109/ICSE55347.2025.00005}, to analyze developers’ use of LLMs on code generation and summarization in their repositories. These two tasks correspond to generating code and comments, respectively. Throughout the paper, we use the term \emph{detectors} to refer to LLM-generated content detectors. We like to stress that unlike previous studies that focus on proposing new detectors such as DetectCodeGPT, our focus is different: we investigate the proportion of code and comments that are likely to be generated by LLMs in the repositories, their characteristics, the differences between company- and community-maintained repositories as well as the bugs \yj{associated with LLM-generated code.} 
Based on the inclusion and \yj{exclusion} criteria, we selected 8 company- and community-maintained repositories from 2021 to 2025: Go-github, Guava, Liquid, Zap, Act, Jadx, Kafka and Pandas. 
We selected the repositories from 2021 because several LLMs, such as GPT-Neo \cite{black2021gpt}, Codex \cite{lu2021codexglue} and Github Copilot \cite{copilot-github}, were released or open-sourced in 2021.
One of the major challenges in performing the detection is the absence of ground truth data (human-labelled LLMs generated content). Since detectors rely on thresholds to classify inputs as likely to be generated by LLMs, it is challenging to determine optimal thresholds without human-labelled data, as these thresholds vary depending on the dataset used. 
\yj{For comments, while previous benchmarks like DetectRL benchmark \cite{wu2024detectrl} offer thresholds for general domains (e.g., Yelp reviews on businesses, human-written stories, etc), these may not accurately reflect the unique linguistic and structural patterns of comments, given that optimal thresholds vary across domains. To ensure that the thresholds are suitable for comments, we utilized the AISE dataset \cite{10.1145/3727582.3728683}, which consists of original repository comments and LLM-generated comments, to derive optimal thresholds for our analysis. Since we cannot guarantee that the original comments were written by humans, we further filtered the AISE dataset to include only comments from files last modified before 2021.}
For code, we applied the default threshold from DetectCodeGPT as it was evaluated on developers' repositories, which is similar to our study. After using detectors to find code and comments that were likely to be LLM-generated, we performed a coding process to categorize them.  We note that a previous study has found that LLM-generated code contains code clones \cite{wu2025empirical}. Thus, we studied the characteristics of these code by analyzing code clones within repositories, across repositories, and in the GPTCloneBench \cite{alam2023gptclonebench}.  We also examined whether comments detected as likely to be LLM-generated exhibit similar properties (e.g., frequent use of AI-related vocabulary such as \emph{aspect} and \emph{capturing}, and having high grammatical accuracy) as summarized in previous work on LLM-based textual output \cite{russell-etal-2025-people}. 
To analyze whether bugs in repositories were \yj{associated with LLM-generated code,}
we used the PreciseBugs dataset \cite{he2023precisebugcollector}. The PreciseBugs dataset was selected because it contains human-labelled bugs introduced after 2021, and several LLMs were released or open-sourced in 2021. This is in contrast to other datasets, such as Defects4J \cite{just2014defects4j} and InferredBugs \cite{jin2023inferfix} where the human-labelled bugs are before 2020 (the pre-LLM era), and they are unlikely to be LLM-generated. \yjnew{In this paper, we refer to the use of detectors to detect code and comments that are likely to be LLM-generated as a detector-based proxy analysis, since the analysis is derived from detector outputs rather than ground-truth annotations of actual LLMs' usage.}

We open-sourced the scripts used in our experiment \footnote{https://github.com/yongyiji/LLM-Generated}.

\yjnew{Based on detector-based proxy analysis, we observed }several interesting phenomena: 1) In active repositories, there is a decreasing trend of code detected as likely to be LLM-generated, while comments detected as likely to be LLM-generated remain stable. Code detected as likely to be LLM-generated primarily appear in test cases, and comments detected as likely to be LLM-generated mainly in Explanation or Meta category; 2) For code detected as likely to be LLM-generated, majority of the repositories have greater than 70\% of file-level code clones and for method-level code clones, only a small percentage was found, with the highest being 34.21\%. When compared with GPTCloneBench, we found that most of the code detected as likely to be generated by LLMs had no code clones.
For comments detected as likely to be generated by LLMs, there is a high proportion of proper punctuation, with an average of around 90\%, which correlates to previous study \cite{russell-etal-2025-people} summarizing the properties of text classified as LLM-generated. However, these comments showed a relatively low percentage of grammatically correct sentences and limited usage of AI-related vocabulary; 3) Company-maintained repositories have a higher percentage of code and comments that are likely to be generated by LLMs, and a higher percentage of code clones found; 4) In code repositories, we found that only a small percentage of the human-labelled bugs (10.79\% and 5.56\% from NVD and OSS-Fuzz, respectively) from the PreciseBugs dataset are likely to be generated by LLMs.

Overall, this paper makes the following contributions:
\begin{itemize}

\item 
We conducted extensive experiments on multiple different repositories across 2021 to 2025 using various detectors to investigate if code and comments are likely to be generated by LLMs in the repositories \yjnew{under a detector-based proxy analysis framework}. \yj{Moreover, we tested a set of thresholds for detecting comments that are likely to be generated by LLMs.}
\item 
We compared company- and community-maintained repositories and found that\yjnew{, based on our detector-based proxy analysis,} the percentage of code that is likely to be generated by LLMs is higher in company-maintained repositories.
\item 
\yjnew{Based on our detector-based proxy analysis,} we analyzed the human-labelled bugs in the repositories and found that only a small percentage of them (10.79\% and 5.56\% from NVD and OSS-Fuzz, respectively) is likely associated with LLM-generated code.
\end{itemize}

The rest of this paper is organized as follows. Section \ref{section:background} provides background for our paper and Section \ref{section:related-work} introduces related work. Section \ref{section:research-design} outlines research questions, data collection and experiment setup. Section \ref{section:results} presents the results of our analysis. This is followed by a discussion in Section \ref{section:discussion} and an analysis of threats and validity in Section \ref{section:threats}. Finally, Section \ref{section:conclusion} concludes and outlines directions for future work.

\section{Background}
\label{section:background}

LLM-generated text is reported to exhibit common traits: it has higher average log probability \cite{gehrmann2019gltr}, it tends to favor high-probability tokens, resulting in lower ranks \cite{gehrmann2019gltr}, and it is more predictable and typically has lower entropy due to concentrated probability distributions \cite{gehrmann2019gltr}.
There are various existing zero-shot methods for detecting LLM-generated content, which can mainly be categorized into three types: statistical-based methods, perturbation-based methods, and model-comparison-based methods \cite{wu2024detectrl, wu-etal-2025-survey}. In this work, we used all the three methods in our experiments as described below.

Statistical-based methods detect LLM-generated text by analyzing token-level statistical features derived from a language model’s probability distribution, such as token log probabilities, token ranks, or entropy. 
Token log probabilities calculate the average log probability of a text,  
token ranks refer to the rank of a token within the model’s probability distribution for a given text and entropy measures the uncertainty of a model’s token prediction distribution. \yj{As an advanced statistical approach, Fast-DetectGPT examines the conditional probability curvature of the log-probability \cite{bao2023fast}. Unlike previous methods that look at a single average value, Fast-DetectGPT analyzes the distribution of alternative token choices at each position.}

Perturbation-based methods, such as DetectGPT and \yj{DetectCodeGPT}, identify LLM-generated text by perturbing the input text and analyzing changes in the model’s output \cite{mitchell2023detectgpt, 10.1109/ICSE55347.2025.00005}. The core idea of perturbation-based methods is that, when text is perturbed or slightly rewritten by alternative LLMs, LLM-generated text tends to be less robust than human-written text. For example, DetectGPT observes that LLM-generated text is highly sensitive to perturbations, resulting in larger drops in log probability compared to human-written text. This is because LLM-generated text is usually more predictable and is constrained by the patterns the model was trained on. Unlike statistical-based methods, which only calculate statistics for the original text, perturbation-based methods calculate the statistics for both original and its perturbed variants.

Model-comparison-based methods, such as Binoculars \cite{10.5555/3692070.3692768}, detect LLM-generated text by comparing the outputs of multiple LLMs. Unlike perturbation-based methods, model-comparison-based methods do not rely on perturbing the input text. Instead, they leverage the observation that LLM-generated text tends to exhibit greater consistency across different models than human-written text.

\section{Related Work}
\label{section:related-work}

\noindent\textbf{Studies on Developers' Use of LLMs}

Previous studies have examined the practices and challenges of using LLMs for developers. Zhang et al. \cite{zhang2023practices} studied discussions on Stack Overflow and GitHub and found that developers often express hesitation and face challenges when incorporating Github Copilot into their workflows. Similarly, Jaworski and Piotrkowski \cite{jaworski2023study} surveyed 42 developers and found that most participants did not want to use Github Copilot. Liang et al. \cite{liang2024large} conducted a survey of 410 developers. They found that the main reasons developers avoid using LLM-based tools are that these tools often fail to generate satisfactory code meeting functional or non-functional requirements, and that developers have difficulty controlling the tools to produce the desired output. At the same time, the study found that developers are motivated to use LLM-based tools because they help reduce keystrokes and can complete programming tasks more quickly.

There were also studies examining the impact of LLMs on developers’ productivity and have reported varying results. Some studies have shown that using LLMs can enhance productivity. For example, Ziegler et al. \cite{ziegler2024measuring} investigated the use of Github Copilot and found that it can offer useful suggestions that guide developers’ progress. Peng et al. \cite{peng2023impact} conducted a controlled experiment to assess the impact of LLM-based tools on professional software development and found that Github Copilot had a statistically significant effect on developers’ productivity. Moreover, Weisz et al. \cite{weisz2025examining} analyzed the impact of using LLMs to assist developers through surveys at a large technology company. They found that while LLM-based tools could increase developers’ productivity, these benefits were not experienced equally by all users. Kuttal et al. \cite{kuttal2021trade} compared human-human and human-agent pair programming and found that LLM-based agents can serve as effective pair programming partners, matching humans in productivity, code quality, and learning experience. On the contrary, some studies found that using LLMs does not always improve work quality. Imai \cite{imai2022github} compared Github Copilot to a human pair programmer and found that although Github Copilot increased productivity, the quality of the code produced was lower than that of human pair programming. 

Our work differs from previous research by analyzing the developers' repositories to check how likely the code and comments are generated by LLMs.

\noindent\textbf{LLM-generated Text Detection} 

In recent years, researchers have been working on detecting LLM-generated text. There are mainly three categories of approaches for LLM-generated text detection: zero-shot methods, watermarking methods and supervised models. 

Zero-shot methods are usually based on the discrepancy between the Log-likelihood and ranking information of human and LLM-generated texts. Gehrmann et al. proposed GLTR \cite{gehrmann2019gltr}, a zero-shot detection method that identifies LLM-generated text by analyzing each token’s rank and Log-likelihood within the probability distribution of a pretrained language model. DetectGPT \cite{mitchell2023detectgpt} detect LLM-generated text by analyzing how small perturbations affect a language model’s log-likelihood of a passage, leveraging the observation that generated text tends to occupy regions of negative curvature in the model’s probability surface. Based on DetectGPT, Bao et al. proposed Fast-DetectGPT \cite{bao2023fast}, which accelerates zero-shot detection by replacing perturbations with a more efficient sampling strategy. Due to their statistical nature, zero-shot methods generally tend to achieve higher detection accuracy on longer passages \cite{lavergne2008detecting, su-etal-2023-detectllm}. Shi et al. proposed DetectCodeGPT \cite{10.1109/ICSE55347.2025.00005} which extends the framework of DetectGPT by perturbing stylistic tokens that capture the distinctive patterns between LLM-generated and human-written code, rather than perturbing arbitrary tokens. DetectCodeGPT is model agnostic and can detect code generated by various LLMs, as it does not rely on training or fine-tuning on any model but instead analyzes the differences between human-written and LLM-generated code \cite{10.1109/ICSE55347.2025.00005}.

Watermarking methods embed token-level markers within generated text, which are invisible to humans, enabling reliable detection of LLM-generated content. These approaches explore the potential of incorporating watermarks into language models to make LLM-generated texts easier to identify. Kirchenbauer et al. \cite{kirchenbauer2023watermark} proposed a statistical watermarking method that divides the vocabulary into green and red token lists based on hash values of preceding n-grams, and softly increases the logits of green tokens during generation to embed a watermark. Based on the work of Kirchenbauer et al., Zhao \cite{zhao2023provable} used a fixed green-red split to propose a more robust watermarking method. However, watermarking approaches depend on model owners, such as OpenAI, to embed the watermark, which limits its broader utilization \cite{10.5555/3692070.3693262}. Moreover, Singh and Zou \cite{singh2023new} found that watermarking method affects text quality, especially in reducing the coherence and depth of the generated responses.

Supervised models are trained on human-labelled datasets to distinguish human-written from LLM-generated text \cite{ippolito2019automatic}. Previous studies have fine-tuned pretrained models to detect synthetic text across various domains, including peer review corpora \cite{bhagat2013paraphrase} and news \cite{uchendu2020authorship, zellers2019defending}. For example, Fagni et al. \cite{fagni2021tweepfake} trained several supervised models to classify LLM-generated content on social media platforms. Similarly, Guo et al. \cite{guo2023close} fine-tuned a RoBERTa-based classifier to distinguish between human-written text and ChatGPT-generated text. In the context of code, Nguyen et al. \cite{nguyen2023snippet} proposed GPTSniffer, which fine-tunes CodeBERT to detect LLM-generated code snippets. However, supervised models trained to detect LLM-generated content may overfit to their training data \cite{uchendu2020authorship}.

We do not consider watermarking methods or supervised classifiers in our analysis because they require training of human-labelled datasets, which we do not possess. 
Moreover, although existing detectors such as DetectCodeGPT have been used to detect code in repositories, they primarily focused on proposing a detector. Contrary to that, our focus is on analyzing the proportion of code and comments that are likely to be generated by LLMs and their characteristics, comparing company- and community-maintained repositories and identifying the bugs \yj{likely associated with LLM-generated code.} 


\noindent\textbf{Bugs in LLMs-generated code} 

Previous studies have found that code generated by LLMs contains bugs. Fan et al. \cite{fan2023automated} analyzed bugs in Codex-generated code and found that such code shares common mistakes with human-written code and exhibits several negative symptoms, including names that indicate incorrect algorithms, duplicated or similar code blocks, and irrelevant helper functions. Dakhel et al. \cite{dakhel2023github} analyzed the quality of code generated by GitHub Copilot and found that some of the generated code contains bugs and is non-reproducible. Liu et al. \cite{liu2024refining} studied the quality of code generated by ChatGPT and summarized the common issues in ChatGPT-generated code. Recently, Tambon et al. \cite{tambon2025bugs} conducted an empirical study on bugs in code generated by LLMs and summarized 10 bug patterns that differ from bugs in human-written code. While the findings of these studies show that code generated by LLMs may contain different kinds of bugs, none of them have examined how many and what bugs are likely to be LLM-generated in repositories. To the best of our knowledge, our study is the first to analyze the proportion and category of bugs in repositories that are likely to be LLM-generated.

\section{Research Design}
\label{section:research-design}

\subsection{Research questions}
The research questions are described as follow:

\begin{itemize}
\item RQ1: How does the proportion of LLM-generated code and comments in repositories, as identified by the detectors, change over time?

According to Google’s DORA report \cite{dora-report}, 90\% of the surveyed developers reportedly use LLMs in their work. However, there is skepticism on the usages of LLMs due to concerns that LLMs generate non-functional code and the unnaturalness of the generated comments \cite{liang2024large, sergeyuk2025using}. This controversy motivated us to investigate whether the repositories contain LLM-generated content, and how the proportion of LLM-generated content changes over time. Moreover, we aim to detect code and comments that are likely to be LLM-generated, given the widespread adoption of LLMs techniques in SE tasks such as code generation and summarization. For each repository, our analysis covers the period from October 2021 to October 2025 because several LLMs, such as GPT-Neo, Codex and Github Copilot, were released or open-sourced in 2021. Starting from this time, developers could use LLMs to generate code and comments. We used the zero-shot based detectors (Binoculars, Log-Likelihood, Entropy, Rank, Log-Rank, LRR, DetectGPT, Fast-DetectGPT and DetectCodeGPT) to detect code and comments that are likely to be generated by LLMs as these detectors analyzed the differences between LLM-generated and human-written content in a LLM agnostic manner. We examined yearly snapshots corresponding to the state of the repository each October. By using detectors to detect content that is likely generated by LLMs at these time points, we can analyze \yjnew{potential LLM usage} evolves within their real-world work. Moreover, we conducted a coding process to categorize the code and comments that were detected as likely to be LLM-generated.

\item RQ2: What are the characteristics of LLM-generated content detected by detectors in repositories?

Previous studies have analyzed the characteristics of LLM-generated content. For natural language text, Russell et al. \cite{russell-etal-2025-people} hired annotators to classify non-fiction English articles into either human-written or LLM-generated, and summarized properties of LLM-generated content to distinguish between LLM-generated and human-written texts. They also reported the frequency of these properties based on annotators’ explanations. We follow their summarized properties to analyze whether the comments detected as likely to be LLM-generated exhibit similar properties. In our experiment, since some properties such as originality are too subjective to be evaluated, we only considered the properties that can be automatically detected, including vocabulary, grammar, punctuation, and spelling.

For code, Wu et al. \cite{wu2025empirical} found that LLM-generated code can contain code clones, and commercial AI code generators produce Type-1 and Type-2 code clones. Because we lack human-labelled ground truth for LLM-generated code, we examine whether code detected as likely to be LLM-generated exhibits such clone patterns. To explore this, we utilized CCFinderX \cite{kamiya2002ccfinder}, a token-based code clone detector, to analyze whether the code detected as likely to be LLM-generated by DetectCodeGPT contain code clones. Moreover, we studied code detected as likely to be LLM-generated have code clone in GPTCloneBench \cite{alam2023gptclonebench} which contains code generated by GPT models. This allows us to find whether code detected as likely to be LLM-generated has code clones with known GPT code.

\item RQ3: How does LLMs usage detected by the detectors differ between community-maintained and company-maintained repositories?

According to statements from large technology companies, LLM-generated content can increase development efficiency, with approximately 20\% to 30\% of code being generated by LLMs in their repositories \cite{google-ceo, microsoft-ceo}. In repositories not maintained by large technology companies, it remains unclear whether the proportion of LLM-generated code is similar. To analyze whether there are differences between company-maintained and community-maintained repositories, we collected data from both sources. We compare whether there is a difference between the two types of repositories in the proportion of code and comments detected as likely to be LLM-generated, as well as in the characteristics of code and comment detected as likely to be LLM-generated.

\item RQ4: \yj{How likely are bugs associated with LLM-generated code?}

Liang et al. \cite{liang2024large} found that developers are unwilling to use LLMs not only because the LLM-generated code does not meet their requirements and it is difficult to control the code generation tools, but also because developers need to spend too much time debugging the code produced by LLMs. It remains unknown \yjnew{whether bugs are associated with code detected as likely to be LLM-generated}. We utilized bugs collected by Ye et al. \cite{he2023precisebugcollector} and applied DetectCodeGPT to analyze which of these human-labelled bugs were likely to be generated by LLM. 


\end{itemize}

\subsection{Project Selection}
\label{section:project-selection}
To ensure the reliability and analytical relevance of our dataset, we established a set of inclusion and exclusion criteria for selecting software repositories from GitHub. The inclusion criteria are defined as follows:

Inclusion Criteria:
\begin{itemize}
    \item Active software development repositories.
    The repository must contain source code files and show at least one commit or merged pull request every month over the past year. This ensures that the project remains active and reflects current software development practices.

    \item Dominant use of mainstream programming languages.
    The primary programming language of the repository must be supported by our LLMs-based detection framework including Java, Python, Go, Ruby, Javascript and PHP. This requirement guarantees compatibility with the tools used in our analysis.

    \item Presence of analyzable textual content.
    The repository must include a sufficient amount of code comments or docstrings, enabling meaningful LLMs-based content analysis and interpretation.

    \item Diverse development entities.
    The collected repositories should include projects maintained by community developers as well as by companies, ensuring that the dataset is representative and reflects the practices and styles of different types of development entities.
\end{itemize}

Exclusion Criteria:
\begin{itemize}
    \item Automatically generated or template-based repositories.
    Repositories whose content is primarily produced by non-LLM tools or explicitly labelled as ``boilerplate,'' ``template,'' or ``generated'' in their names or descriptions are excluded.

    \item Forked repositories.
    To avoid redundancy, only original repositories are analyzed. Exceptions are made if a forked repository demonstrates significantly higher activity than its source project.

    \item   Non-software repositories.  
    Repositories that primarily contain datasets, configuration files, or textual materials with minimal or no source code are excluded from the analysis.


\end{itemize}

After applying these criteria, we obtained a refined set of active and analyzable software repositories that align with the goals of our study. The selected repositories cover a wide range of application types (foundational libraries, developer tools, distributed systems, data processing frameworks and template engines) and are written in different programming languages (Python, Java, Go, and Ruby).


\subsection{Data Collection}

According to the inclusion and exclusion criteria, we selected eight repositories in GitHub \cite{GitHub_GraphQL_API}, four company-maintained repositories and four community-maintained projects repositories. For the selected repositories, we fixed our analysis on a specific version snapshot, recording the commit hash, timestamp, and license. The time period spans from October 2021 to October 2025, with data collected yearly. We analyzed the repository starting from 2021 because major breakthroughs in LLMs occurred in 2021 \cite{10.1145/3586030} -- for example, the release of Codex \cite{chen2021evaluating} and the launch of GitHub Copilot \cite{copilot-github} enabled developers to generate code and comments using LLMs \cite{10.1145/3586030}. We employed zero-shot detectors to detect code and comments that are likely to be generated by LLMs, without restriction to any specific LLM \cite{mitchell2023detectgpt, 10.1109/ICSE55347.2025.00005}.

For all repositories, we applied the same data preprocessing procedure to extract the file content and we classified them into two categories: code, and comment. Since LLMs have been widely used in the field of SE for code generation and code summarization \cite{ciniselli2021empirical, ahmed2022few, shojaee2023execution, sun2024source, lu2021codexglue}, we focus on detecting LLM-generated code and comments. We analyzed source files that DetectCodeGPT can analyze, including those written in Java, JavaScript, Python, Go, Ruby, and PHP. 

\subsection{Experiment Setup}

All experiments are conducted on 10 NVIDIA A100 GPUs with 40GB memory. For the hyperparameters used in DetectCodeGPT, we followed previous work \cite{10.1109/ICSE55347.2025.00005} by setting the span length to 2 and masking 50\% of the words during text perturbation. The perturbation type was random-insert-space+newline. For the hyperparameters used in Binoculars, Log-Likelihood, Entropy, Rank, Log-Rank, LRR, DetectGPT and Fast-DetectGPT, we used the same settings as those used in DetectRL \cite{wu2024detectrl}.

\begin{figure*}
    \centering
    \includegraphics[width=1\linewidth]{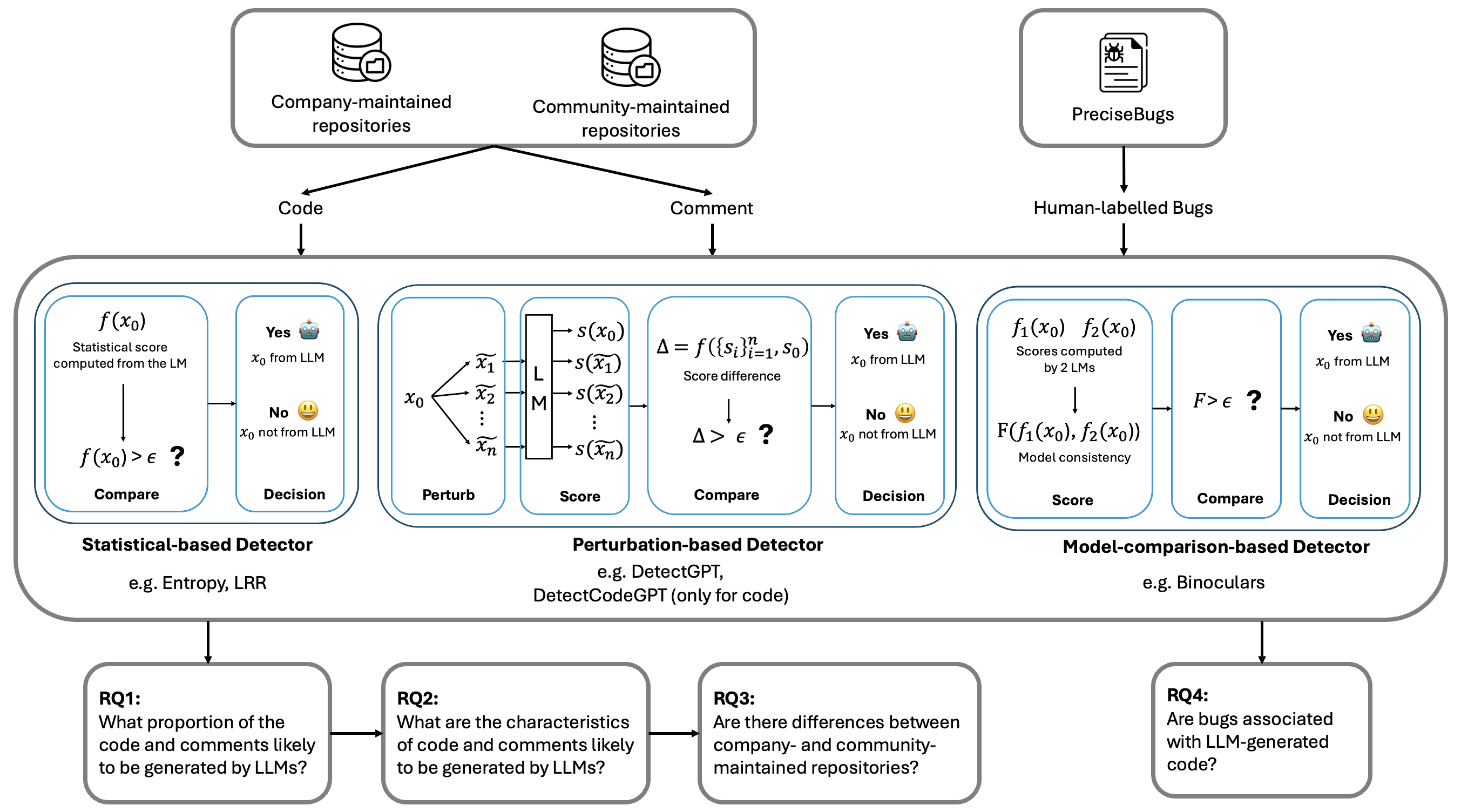}
    \caption{Overall framework of our analysis. The framework illustrates how detectors are applied to identify LLM-generated code and comments in repositories.}
    \label{fig:framework}
\end{figure*}

Figure \ref{fig:framework} shows the overview of the workflow in this study. We apply detectors to identify code and comments that are likely to be generated by LLMs on both company- and community-maintained repositories. The outputs from these detectors are then used to address the four research questions.

\begin{itemize}

\item RQ1: How does the proportion of LLM-generated code and comments in repositories, as identified by the detectors, change over time?

In the experiment, we employed different zero-shot detectors, including Binoculars, Log-Likelihood, Entropy, Rank, Log-Rank, LRR, DetectGPT, Fast-DetectGPT and DetectCodeGPT. To detect comments written in natural language, we used Binoculars, Log-Likelihood, Entropy, Rank, Log-Rank, LRR, DetectGPT, and Fast-DetectGPT. For detecting code, we used DetectCodeGPT, which is specifically designed to identify whether code is LLM-generated or not.

For each detector, the optimal decision threshold depends on the dataset and domain. Existing non-code based detectors do not offer a predefined threshold to determine whether a given input is LLM-generated, as the optimal threshold varies across datasets and is typically derived using the Area Under the Receiver Operating Characteristic Curve (AUROC) \cite{hanley1982meaning}. AUROC metric is widely used for evaluating zero-shot methods \cite{mitchell2023detectgpt}, as it considers both true positive and false positive rates across different decision thresholds. However, in our experiment, there is no ground-truth labels for our dataset, as all data were collected from public repositories. Thus, we were unable to compute the AUROC metric and to find an optimal threshold \yj{for each repository}. 
\yj{Moreover, Wu et al. \cite{wu2024detectrl} proposed DetectRL benchmark which evaluates a set of detectors, including Binoculars, Log-Likelihood, Entropy, Rank, Log-Rank, LRR, DetectGPT and Fast-DetectGPT. It performs different domain evaluation and assesses the generalization of detectors by applying thresholds obtained from one domain to other domains. The DetectRL provided four benchmark datasets: the arXiv dataset for scientific papers, the XSum dataset for news articles, WritingPrompts (WP) for creative stories, and Yelp Reviews for social reviews. However, optimal thresholds vary significantly across domains. The threshold set for each detector in DetectRL was shown in Table \ref{tab:detector_scores}. }
\yj{Given that code comments possess unique structural and linguistic patterns distinct from news or social reviews, directly applying the non-SE thresholds provided by DetectRL could lead to inaccurate results. To ensure that the thresholds are suitable for the SE domain, we used the AISE dataset \cite{10.1145/3727582.3728683} to derive thresholds for code comments. The AISE dataset was proposed by Katzy et al. \cite{10.1145/3727582.3728683}, who conducted an empirical study on LLM-generated code comments and created a dataset consisting of original comments and comments generated by LLMs. They first compiled a list of common words and used GitHub to collect files containing these words. For each file, comments were extracted, and CodeGemma, CodeLlama, CodeQwen, GraniteCode, and StarCoder were used to generate corresponding comments. Since it is unclear whether the original comments were written by humans or generated by LLMs, we filtered out files modified after 2021 to ensure that the retained comments were written by humans. }


\begin{table}[htbp]
\centering
\caption{Detector threshold across four datasets \yj{in DetectRL benchmark}: ArXiv, XSum, WP, and Yelp. }
\footnotesize
\label{tab:detector_scores}
\begin{tabular}{lcccc}
\toprule
\textbf{Detector} & \textbf{ArXiv} & \textbf{XSum} & \textbf{WP} & \textbf{Yelp} \\ 
\midrule
Binoculars & -0.92 & -0.92 & -0.91 & -0.93 \\ 
Log-Likelihood & -2.24 & -2.25 & -2.64 & -2.57\\ 
Entropy & 2.96 & 2.95 & 3.47 & 3.79 \\ 
Rank & -52.51 & -17.72 & -46.88 & -39.83\\ 
Log-Rank & -1.12 & -1.09 & -1.28 & -1.328 \\ 
LRR & 2.07 & 2.11 & 2.03 & 2.03\\ 
DetectGPT & 1.03 & 0.94 & 0.54 & 0.50 \\ 
Fast-DetectGPT & 5.48 & 5.16 & 4.64 & 4.19 \\ 
\bottomrule
\end{tabular}
\end{table}

\yj{However, there is a difference in token length between the AISE dataset and the comments in our dataset, as shown in Table \ref{tab:length}. While the AISE dataset primarily consists of shorter comments, real-world repositories often contain much longer comments. Previous study has shown that detectors exhibit a performance degradation when processing short text \cite{chakraborty2023possibilities}. Therefore, we filtered the AISE dataset to exclude comments containing fewer than 10 tokens and computed the AUROC metric to determine the threshold. The threshold for SE domain is shown in Table \ref{tab:threshold_se}. Although detectors are primarily designed for general domains, comments are written in natural language. Therefore, zero-shot detectors developed for natural language are inherently applicable to the SE domain. For DetectCodeGPT, we used the dataset provided in DetectCodeGPT \cite{10.1109/ICSE55347.2025.00005} to obtain the threshold, which was then applied in our experiment.} \yjnew{The threshold was derived from open-source Github projects aross multiple programming languages. This is consistent with our experiment.}

\begin{table}[htbp]
\centering
\caption{Distribution of comment lengths in the AISE dataset and our repository dataset.}
\footnotesize
\label{tab:length}
\begin{tabular}{lcc}
\toprule
\multirow{2}{*}{\textbf{Length}} & \multicolumn{2}{c}{\textbf{Distribution}} \\
& \textbf{Repos} & \textbf{AISE} \\
\midrule
0-10 & 7.01\% & 19.16\%\\
10-20 & 10.53\% & 23.82\%\\
20-50 & 17.51\% & 31.19\%\\
50-100 & 13.49\% & 14.81\%\\
100-500 & 33.55\% & 11.03\%\\
500+ & 17.91\% & 0.00\%\\
\bottomrule
\end{tabular}
\end{table}

\begin{table}[htbp]
\centering
\caption{Detector thresholds for comments in the software engineering domain, derived from the AISE dataset after filtering out comments with fewer than 10 tokens.}
\footnotesize
\label{tab:threshold_se}
\begin{tabular}{lcccccc}
\toprule
\textbf{Detector} & \textbf{Threshold} & \textbf{F1} & \textbf{TP} & \textbf{FP} & \textbf{TN} &\textbf{FN}\\
\midrule
Binoculars & -1.00 & 0.83 & 553 & 71 & 154 & 149\\
Log-Likelihood & -3.20 & 0.58 & 303 & 36 & 189 & 299\\
Entropy & 2.12 & 0.61 & 339 & 62 & 163 & 363\\
Rank & -163.38 & 0.78 & 516 & 108 & 117 & 186\\
Log-Rank & -1.13 & 0.72 & 437 & 68 & 157 & 286\\
LRR & 1.83 & 0.70 & 416 & 68 & 157 & 286\\
DetectGPT & 0.04 & 0.71 & 457 & 127 & 98 & 245\\
Fast-DetectGPT & 1.18 & 0.79 & 486 & 35 & 190 & 216\\
\bottomrule
\end{tabular}
\end{table}




After using detectors to detect code and comments that are likely to be LLM-generated, we conducted a coding process to identify the categories these code and comments appeared in. For comments, we followed the previous work \cite{padioleau2009listening} which summarized the comment classifications. Padioleau et al. \cite{padioleau2009listening} classified comments into different angles including ``what", ``who", ``when", ``where". We only applied the ``what" dimension from Padioleau et al.'s work because we are concerned on the purpose of the comments detected as likely to be LLM-generated. The codebook for labelling comments is shown in Table \ref{tab:codebook}. As Padioleau et al.’s work studied comments from open-source software written in C, we adapt their categories to support other programming languages. The categories, Code Relationship, PastFuture, Meta, and Explanation, retain the same definitions as in Padioleau et al.’s work, whereas the categories, Type and Interface, are adapted to broader interpretations. The original Type category included C specific subcategories such as Unit, IntRange, and BitsBytes. We refine this category to capture value constraints, units, ranges, formats, and conceptual types that appear across languages. Similarly, the Interface category is broadened into a more general interface contract that captures how functions or modules should be used. For each repository, we considered all comments that were detected as likely to be LLM-generated by any detector between 2021 and 2025, and randomly sampled 370 comments from this combined set. The sampled dataset satisfy a 95\% confidence level and a 5\% margin of error \cite{yang2023demystifying}. Two authors independently coded the first 100 comments in the sampled dataset. For the initial coding, the Cohen's kappa score is 0.63 \cite{byrt1996good}. The two authors then discussed their disagreements and reached an agreement on the definitions. Subsequently, they independently coded the remaining sampled comments, achieving a Cohen’s Kappa score of 0.76 \cite{byrt1996good}. For any inconsistencies, two authors discussed and reached an agreement.

For code, we created a codebook to categorize code, since previous studies did not provide such categorization for code. For each repository, we randomly sample 370 code that is detected as likely to be LLM-generated between 2021 and 2025. The sample size for each repository satisfies a 95\% confidence level and a 5\% margin of error \cite{yang2023demystifying}. Following that, two authors independently analyzed the first 370 samples to create an initial list of categories (codebook). They then annotated the remaining sampled code based on the initial codebook. During the annotation process, when code was encountered that do not fit into the initial codebook, the authors met to discuss and refine the codebook. Through iterative discussions, two authors collaboratively developed a finalized codebook shown in Table \ref{tab:codebook} and independently coded all the sampled code, with a Cohen’s Kappa score of 0.78 \cite{byrt1996good}. Finally, the authors discussed any disagreements to reach a consensus.

\begin{table}[!ht]
  \centering
  \footnotesize
  \caption{Definitions of categories used to classify code or comments detected as likely to be LLM-generated. Categories listed in the upper section refer to code detected as likely to be LLM-generated, whereas those listed in the lower section refer to comments detected as likely to be LLM-generated.
 }
  \label{tab:codebook}
    \begin{tabular}{lp{50mm}}
    \toprule
    \textbf{Category}
      & \textbf{Description}\\

    \midrule
    \textbf{Core logic} & Contains the core rules, business logic, and main algorithms that implement the system’s essential functionality.\\
    \textbf{Interfaces} & Defines the programming contracts, public APIs, and extension points for interacting with the core logic.\\
    \textbf{Common} & Provides shared infrastructure, utilities, and technical implementations.\\
    \textbf{Tests} & Includes test cases used for verification and validation.\\
    \midrule
    \textbf{Type} & Specifies value meanings or constraints.\\
    \textbf{Interface} & Describes the behavioral contract of functions, methods, or modules.\\
    \textbf{Code Relationship} & Specifies some code relationships.\\
    \textbf{PastFuture} & Describes code evolution aspects, such as past changes, current issues, or future tasks.\\
    \textbf{Meta} & Provides authorship, licensing, or other non-behavioral metadata.\\
    \textbf{Explanation} & Comments not covered by the other five categories.\\
    
    \bottomrule
  \end{tabular}

\end{table}

\item RQ2: What are the characteristics of LLM-generated content detected by detectors in repositories?

To analyze the code detected as likely to be LLM-generated, we employed CCFinderX, a token-based clone detection tool designed for identifying similar code fragments and is widely used in the SE field \cite{kamiya2002ccfinder}. CCFinderX detects code clones by tokenizing code into token sequences and identifying repeated or structurally equivalent subsequences across files. During preprocessing, the tool applies normalization to the code, which enables the detection of both Type-1 and Type-2 clones. In Type-2 clones, identifiers and literals may differ, but the overall syntactic structure remains the same. In our experiment, we conducted code clone detection within each repository to identify intra-repository code clones and between each repository and the other seven repositories to identify inter-repository clones. Moreover, we conducted a code clone analysis between code detected as likely to be LLM-generated and the GPTCloneBench dataset \cite{alam2023gptclonebench}. The GPTCloneBench \cite{alam2023gptclonebench} is a benchmark dataset for GPT-generated semantic and cross-language code clones, validated through both manual and automated verification. The GPTCloneBench dataset includes code clones across four programming languages: Python, Java, C, and C\#. As the selected repositories (Guava, Zap, Jadx, Kafka, and Pandas) are primarily written in Java and Python, our comparison with GPTCloneBench was limited to these repositories.

For comments, we analyzed the characteristics of comments detected as likely to be LLM-generated applying the guidelines proposed by Russell et al. \cite{russell-etal-2025-people}, which provide a guide to distinguish LLM-generated text from human writing. Since some of the criteria are subjective like originality and tone, we only applied those that can be assessed automatically, including detecting words frequently generated by AI and checking spelling, punctuation and grammar. Following the work of Russell et al. \cite{russell-etal-2025-people}, the AI-related words include nouns, verbs, adjectives, and adverbs. To evaluate spelling, punctuation and grammar, we applied automated text analysis at the sentence level using the language\_tool\_python library \cite{language_tool_python}, which provides sentence-level linguistic error detection.

\item RQ3: How does LLMs usage detected by the detectors differ between community-maintained and company-maintained repositories?

Following the inclusion and exclusion criteria described in Section \ref{section:project-selection}, we selected the repositories shown below, which contain both company- and community-maintained repositories.

The company-maintained repositories are 
Go-github \cite{go-github}, Guava \cite{guava}, Liquid \cite{liquid}
and Zap \cite{zap}. These repositories are maintained by major technology companies, including Google, Meta, Uber, and Shopify.

The community-maintained repositories are Act \cite{act}, 
Jadx \cite{jadx}, Kafka \cite{kafka}, and Pandas\cite{pandas}. 
These repositories are primarily maintained by the open-source communities, rather than by large technology companies.

For all the repositories, we followed the same process described in the RQ1 experiment design to detect comments, and code separately, and to compare differences in code and comments detected as likely to be LLM-generated between company-maintained and community-maintained repositories.

\item RQ4: \yj{How likely are bugs associated with LLM-generated code?} 

To answer this question, we detect the human-labelled bugs to see if they were LLM-generated. 
However, as our dataset does not contain human-labelled bugs, we used a bug dataset, PreciseBugs, curated by Ye et al. \cite{he2023precisebugcollector}, containing 1,057,818 bugs from 2,968 open-source repositories and it includes multiple programming languages, such as C/C++, Rust, Go, Python, and Java/JVM. We focused on the Rust, Go, Python, and Java bugs due to the limitation of DetectCodeGPT, which cannot detect code written in C and C++. In the PreciseBugs dataset, there are three sources: two are human-labelled bugs (NVD and OSS-Fuzz) found in repositories, and one is a synthesized version. NVD (National Vulnerability Database) stores the standardized vulnerability reports, and OSS-Fuzz discovers bugs in open-source software. 
We only considered the human-labelled bugs, as we aimed to analyze whether the bugs in the repositories are likely to be generated by LLMs. In addition, we limited our analysis to the human-labelled bugs generated after October 2021, which aligns with the period covered in our study. \yj{Moreover, we define the unit of analysis for this study as the entire source code file containing the labelled bug.}

\end{itemize}

\section{Results}
\label{section:results}

\begin{figure}
    \centering
    \includegraphics[width=1\linewidth]{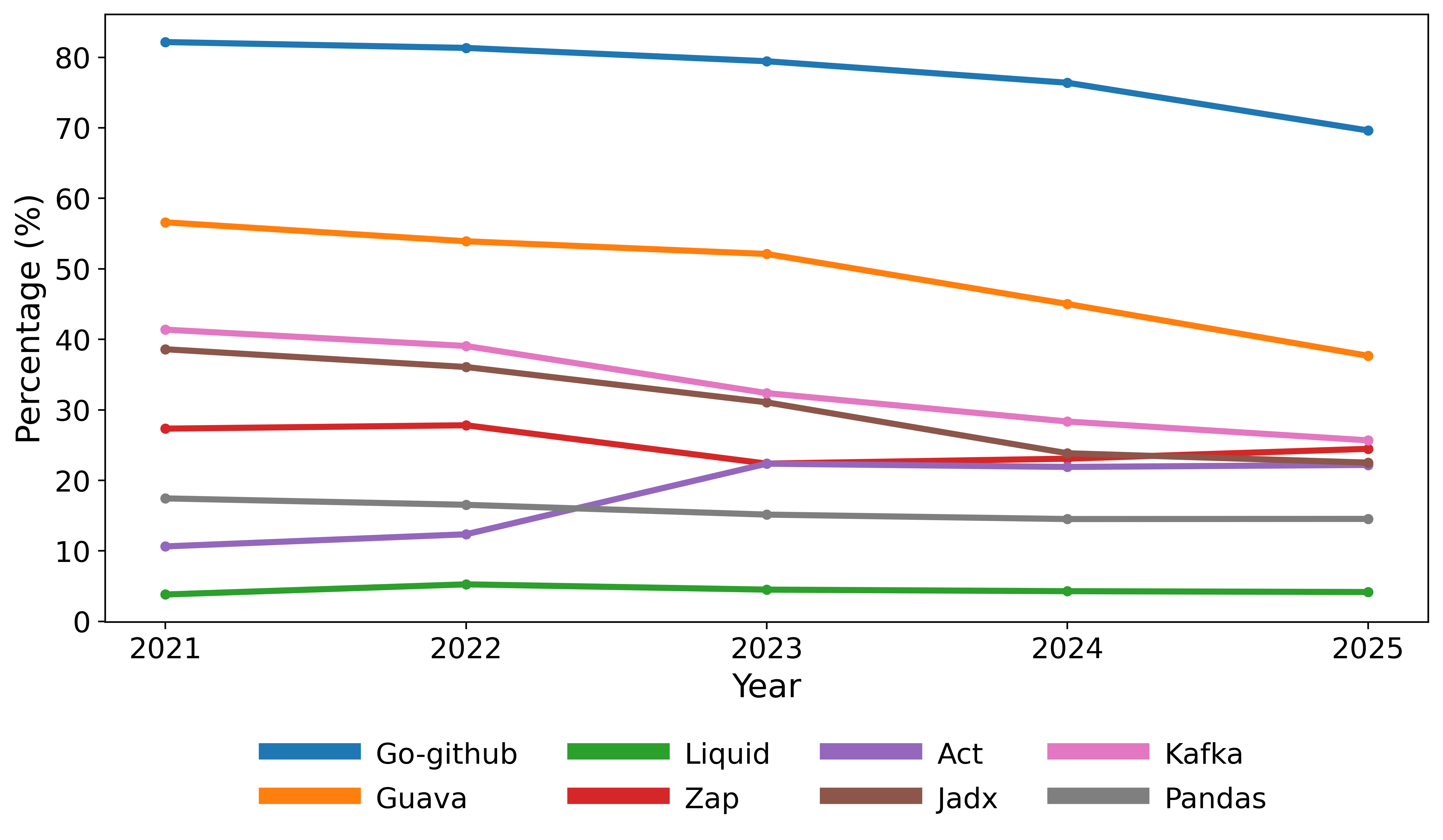}
    \caption{The percentage of code detected as likely to be LLM-generated in repositories detected by DetectCodeGPT. Each line represents one repository. The repositories, Go-github, Guava, Liquid, and Zap represent the company-maintained repositories, while the repositories, Act, Jadx, Kafka, and Pandas represent the community-maintained repositories.}
    \label{fig:detectcodegpt_result}
\end{figure}


\begin{figure*}[!t]
    \centering
    \subfigure[Company-maintained repositories]{
        \begin{minipage}[b]{0.48\textwidth}
            \centering
            \includegraphics[width=\textwidth]{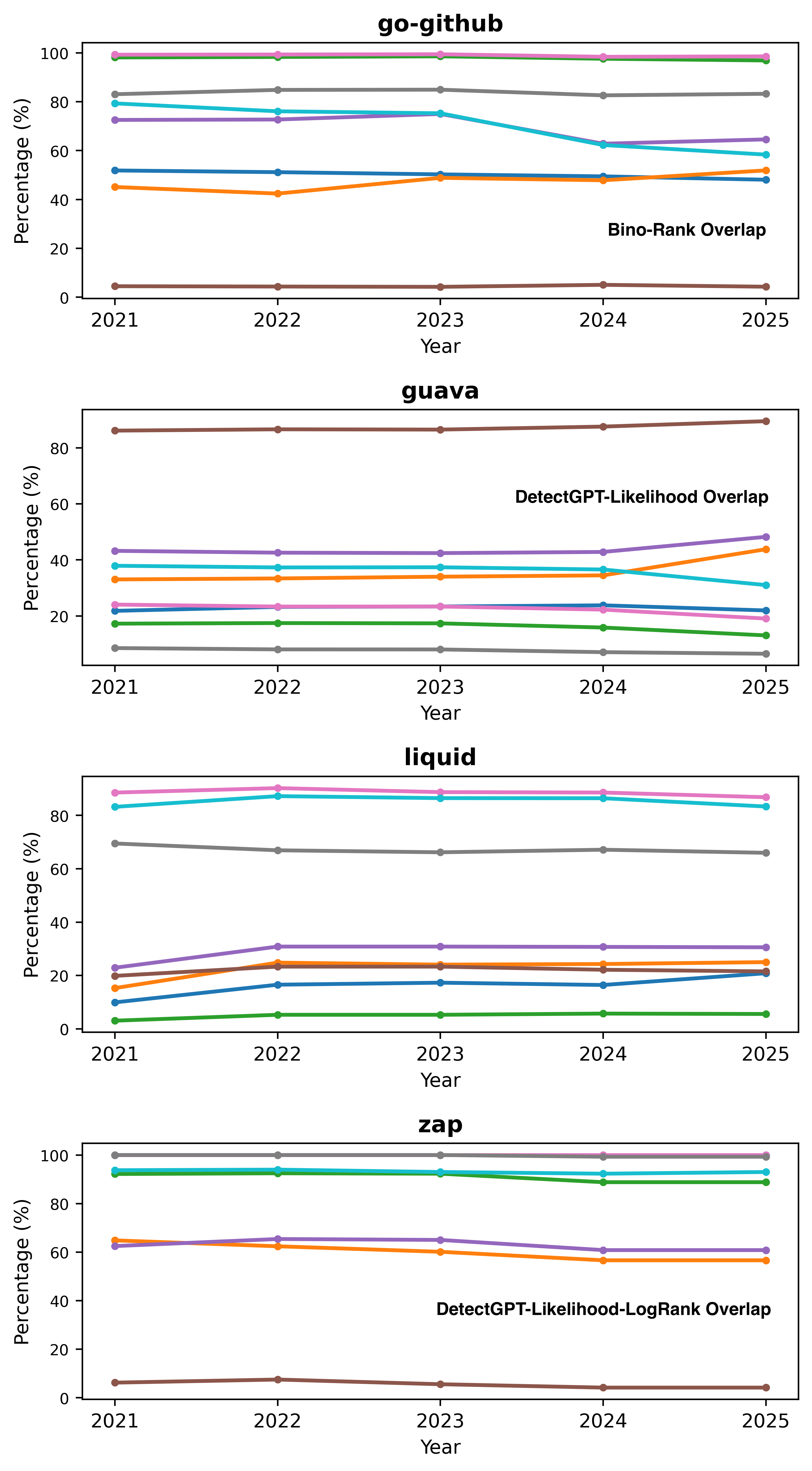}
            \vspace{-1em}
            \label{fig:company_avg}
        \end{minipage}
    }
    \hfill
    \subfigure[Community-maintained repositories]{
        \begin{minipage}[b]{0.48\textwidth}
            \centering
            \includegraphics[width=\textwidth]{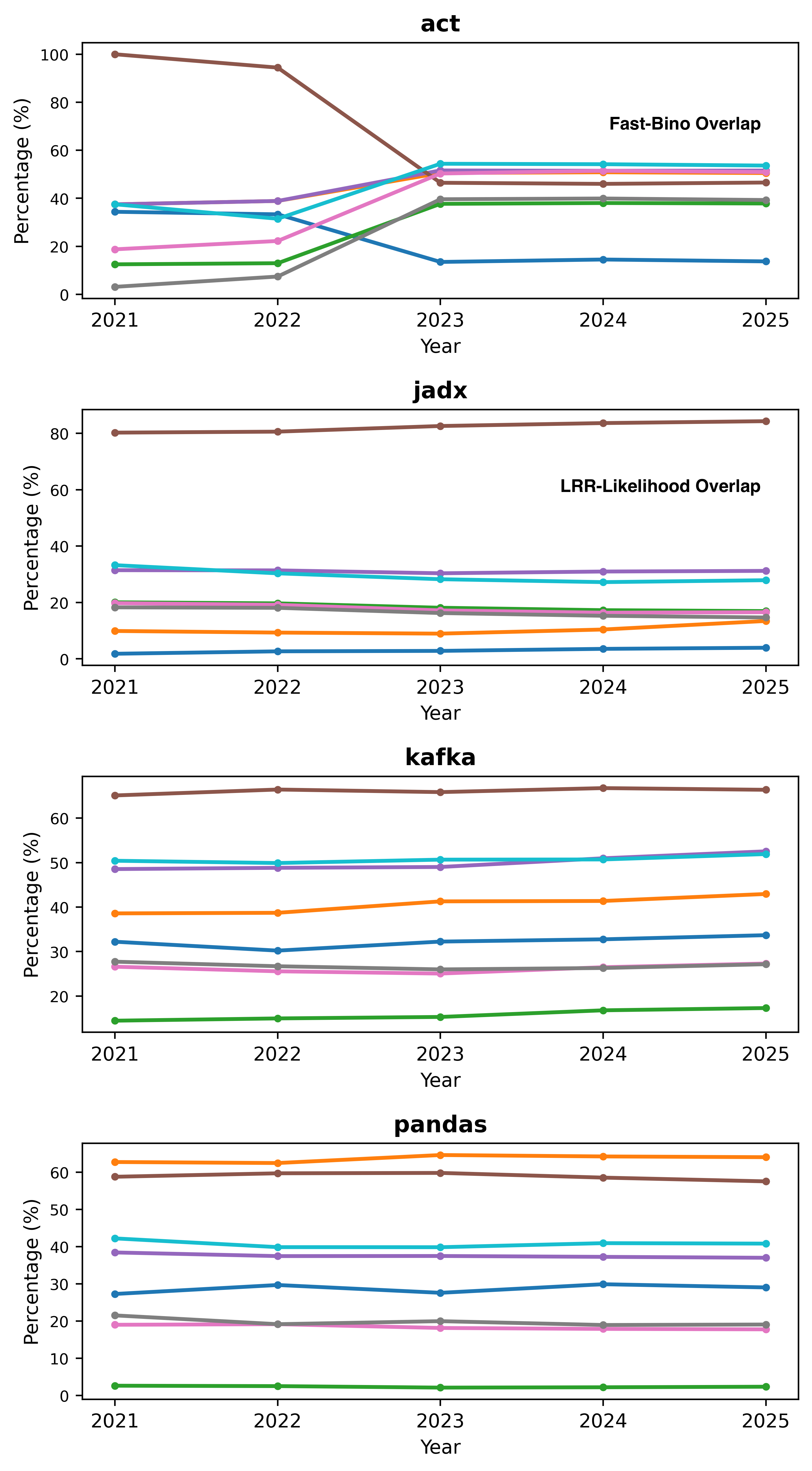}
            \vspace{-1em}
            \label{fig:community_avg}
        \end{minipage}
    }

    \begin{minipage}{0.6\textwidth}
        \centering
        \includegraphics[width=\textwidth]{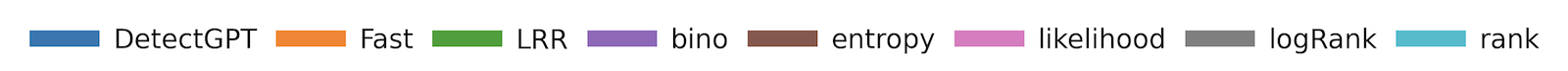}
    \end{minipage}

    \caption{The percentage of comments detected as likely to be LLM-generated in each repository over time, 
    \yj{using the threshold obtained from AISE dataset.}
    The different colored lines represent the different detectors: the blue line represents DetectGPT, the orange line represents Fast-DetectGPT, the green line represents LRR, the purple line represents Binoculars, the brown line represents Entropy, the pink line represents Log-likelihood, the gray line represents Log-Rank, and the sky-blue line represents \yj{Rank}.
    Figure \ref{fig:company_avg} shows the percentage change in likely to be LLM-generated comments from the company-maintained repositories between 2021 and 2025, while Figure \ref{fig:community_avg} shows the percentage change in likely to be LLM-generated comments from the community-maintained repositories between 2021 and 2025.
    The overlapped lines are mentioned in wordings in the plots.
    \yj{The following detector lines overlap in the corresponding repositories: in Go-github: Binoculars and Rank; in Guava: DetectGPT and Log-likelihood; in Zap: DetectGPT, Log-likelihood and Log-Rank; in Act: Fast-DetectGPT and Binoculars; in Jadx: LRR and Log-Likelihood.}
    }
    \label{fig:comment_avg}
\end{figure*}

\subsection{RQ1: How does the proportion of LLM-generated code and comments in repositories, as identified by the detectors, change over time? }

Figure \ref{fig:detectcodegpt_result}, \ref{fig:comment_avg} show the proportions of likely to be LLM-generated content in each repository detected by each detector. Overall, the proportion of code detected as \yjnew{likely to be} LLM-generated by the detectors decreased over time. As shown in Figure \ref{fig:detectcodegpt_result}, out of the eight code repositories, 50\% showed a decrease in the proportion of code detected as likely to be LLM-generated. For example, the Go-github and Guava repositories showed a substantial decrease in the proportion of code detected as likely to be LLM-generated, dropping from 82.16\% to 69.62\% and 56.6\% to 37.67\% respectively, between 2021 and 2025. In the Kafka repository, the proportion of code detected as likely to be LLM-generated decreased from 41.38\% to 25.68\%. 12.5\% of the repositories showed an increasing trend in the proportion of code detected as likely to be LLM-generated. The Act repository showed a substantial increase from 10.64\% to 22.19\%. For the remaining repositories, the changes were relatively small. The proportion of code detected as likely to be LLM-generated in the Liquid repository showed a slight increase, while the Zap repository exhibited a slight decrease. 
\yjnew{However, the relatively high proportions detected in earlier years (2021--2023) should not be interpreted as direct evidence of actual LLM usage. Possible explanations for these early-year signals are discussed further in Section \ref{section:discussion}. Therefore, the early-year results should be interpreted primarily as baseline reference points for relative temporal comparison rather than definitive indicators of real-world LLM adoption. In addition, the threshold sensitivity analysis discussed in Section \ref{section:discussion} suggests that the overall trends remain relatively consistent under threshold variations.}

For comments, many of the repositories exhibited a relatively stable proportion of comments detected as likely to be LLM-generated throughout the analysis period, as shown in Figure \ref{fig:comment_avg}. Different detectors showed varying proportions, but similar trends were observed within each repository. The Act repository showed fluctuations in the proportion of comment detected as likely to be LLM-generated before 2023. Among the eight detectors, five detectors, including Log-Rank, Log-likelihood, Rank, LRR, and Binoculars showed an increase, while the other three showed a decrease. After 2023, the proportions of comments detected as likely to be LLM-generated became stable for each detector, with negligible changes in 2024 and 2025. The other repositories showed a stable proportion of comments detected as likely to be LLM-generated. \yjnew{In addition, the threshold sensitivity analysis discussed in Section \ref{section:discussion} indicates that these overall patterns remain relatively stable under threshold variations.}

Table \ref{tab:coding-process} shows the results of the coding process used to categorize the code and comments detected as likely to be LLM-generated. \yj{We filter out code from third-party dependencies; auto-generated artifacts, such as OpenAPI stubs and minified bundles; standardized library files containing external copyright notices (e.g., Microsoft Corp.)} The majority of the code detected as likely to be LLM-generated is in the Tests category, with \yj{62.5\%} of the repositories showing the highest proportion as likely to be LLM-generated in this category. \yj{For instance, the Go-github and Pandas repositories also showed high proportions of Tests category  detected as likely to be LLM-generated, with 61.6\% and 67.5\%, respectively.} Compared with the Tests category, the other categories contained a smaller proportion of code detected as likely to be LLM-generated. For comments detected as likely to be LLM-generated, the Meta and Explanation categories account for a large proportion of comments, while the Type, Interface, Code relationship and PastFuture made up a relatively small proportion. For example, in Go-github repository, 97\% of the comments detected as likely to be LLM-generated were in the Meta and Explanation categories.

\begin{table*}[!ht]
  \centering
  \footnotesize
  \caption{The proportion of the code and comments detected as likely to be LLM-generated within each category. Categories listed in the upper section refer to code detected as likely to be LLM-generated, whereas those listed in the lower section refer to comments detected as likely to be LLM-generated. The first four columns represent the company-maintained repositories, while the last four columns represent the community-maintained repositories.
 }
  \label{tab:coding-process}
    \begin{tabular}{lcccc|cccc}
    \toprule
    \textbf{Category}
      & \textbf{Go-github}
      & \textbf{Guava}
      & \textbf{Liquid}
      & \textbf{Zap}
      & \textbf{Act}
      & \textbf{Jadx}
      & \textbf{Kafka}
      & \textbf{Pandas}\\

    \midrule
    \textbf{Core logic} & 0.5\% & 26.1\% & 40.0\% & 47.3\% & 25.8\% & 21.2\% & 17.1\% & 20.2\% \\
    \textbf{Interfaces} & 37.4\% & 14.3\% & 0.0\% & 7.1\% & 25.8\% & 17.2\% & 32.7\% & 4.8\% \\
    \textbf{Common} & 0.5\% & 26.7\% & 16.7\% & 12.5\% & 30.9\% & 12.1\% & 25.1\% & 7.5\% \\
    \textbf{Tests} & 61.6\% & 32.9\% & 43.3\% & 33.1\% & 17.5\% & 49.5\% & 25.1\% & 67.5\% \\
    \midrule
    \textbf{Type} & 0.0\% & 3.0\% & 6.5\% & 0.0\% & 1.0\% & 0.0\% & 0.0\% & 2.0\% \\
    \textbf{Interface} & 0.0\% & 3.0\% & 1.0\% & 0.0\% & 3.0\% & 0.0\% & 0.0\% & 1.5\% \\
    \textbf{Code Relationship} & 1.0\% & 1.0\% & 1.0\% & 0.0\% & 3.5\% & 5.5\% & 5.5\% & 0.0\% \\
    \textbf{PastFuture} & 2.0\% & 26.0\% & 3.0\% & 0.0\% & 7.0\% & 13.5\% & 1.5\% & 43.5\% \\
    \textbf{Meta} & 34.0\% & 0.0\% & 74.0\% & 71.5\% & 60.0\% & 31.5\% & 11.5\% & 1.0\% \\
    \textbf{Explanation} & 63.0\% & 67.0\% & 14.5\% & 28.5\% & 22.5\% & 49.5\% & 81.5\% & 52.0\% \\
    
    \bottomrule
  \end{tabular}

\end{table*}

\begin{tcolorbox}[colback=black!5!white, colframe=black, title=Answer to RQ1, fonttitle=\bfseries]
Code detected as likely to be LLM-generated decreased for majority of the repositories, while comments detected as likely to be LLM-generated remained relatively stable, with negligible changes across the years. Code detected as likely to be LLM-generated were primarily in the Tests category, while comments detected as likely to be LLM-generated were mostly found in Meta or Explanation category.
\end{tcolorbox}

\subsection{RQ2: What are the characteristics of LLM-generated content detected by detectors in repositories?}

\begin{table}[!ht]
  \centering
  \footnotesize
  \caption{The percentage of LLM-generated code detected by DetectCodeGPT that contains code clones. Intra represents comparisons within each repository. Inter refers to comparisons between a given repository and the other nine repositories at the file level. We define file-level clones as clones identified by comparing entire files containing code detected as likely to be LLM-generated. Method-level clones are identified by splitting those files into individual methods and applying CCFinderX to detect clones at the method level. 
  Company-maintained repositories are shown above, and community-maintained repositories are shown below.
 }
  \label{tab:code-clone}
    \begin{tabular}{ccccc}
    \toprule
    \multirow{2}{*}{\textbf{Repository}} & 
    \multicolumn{2}{c}{\textbf{Intra}} &
    \multicolumn{2}{c}{\textbf{Inter}} \\
    & \textbf{File} & \textbf{Method} & \textbf{File} & \textbf{Method} \\

    \midrule
    \textbf{Go-github} & 97.89\% & 34.98\% & 0.00\% & 0.00\%\\
    \textbf{Guava} & 91.56\% & 26.30\% & 0.00\% & 0.00\%\\
    \textbf{Liquid} & 100\% & 34.21\% & 0.00\% & 0.00\%\\
    \textbf{Zap} & 90.5\% & 33.33\% & 0.00\% & 0.00\%\\

    \midrule
    \textbf{Act} & 93.05\% & 22.81\% & 0.00\% & 0.00\%\\
    \textbf{Jadx} & 85.98\% & 14.55\% & 0.00\% & 0.00\%\\
    \textbf{Kafka} & 77.04\% & 23.16\% & 0.00\% & 0.00\%\\
    \textbf{Pandas} & 45.32\% & 26.29\% & 0.00\% & 0.00\%\\
    
    \bottomrule
  \end{tabular}

\end{table}
Table \ref{tab:code-clone} presented the code clone results for code detected as likely to be LLM-generated, analyzed within individual repositories and across other repositories. \yjnew{A large proportion of code detected as likely to be LLM-generated exhibits file-level clones within their own repository}, with intra-repository file-level clone percentages exceeding 70\% in most cases, except for Pandas (45.32\%). For the Go-github, Guava, Liquid, Zap and Act repositories, the proportion of code detected as likely to be LLM-generated approaches 100\%. For intra-repository method-level clones, the percentage is relatively low, ranging from around 10\% to 35\%. In the inter-repository clone analysis, there is no code clones detected, and we hypothesized that this is due to the analyzed repositories written in different programming languages and that they have different code structures. Table \ref{tab:gpt-clone} showed the code clone results for the code detected as likely to be LLM-generated when compare with GPTCloneBench. Only a small proportion of the code detected as likely to be LLM-generated contains clones in GPTCloneBench. In the Zap and Pandas repositories, no code detected as likely to be LLM-generated had code clones in GPTCloneBench, while in the Guava, Jadx, and Kafka repositories, the proportion of likely to be LLM-generated code with clones in GPTCloneBench was mostly lower than 1\%, with the highest being 3.09\%. This \yjnew{indicates} that developers may not directly use GPT outputs into repositories.

Tables \ref{tab:criteria_company_avg} and \ref{tab:criteria_community_avg} showed the characteristics of likely to be LLM-generated comments, as detected by the detectors using the average thresholds. The percentage of AI-related words used in in the SE domain is relatively low, possibly because the AI vocabulary proposed by Russell et al. \cite{russell-etal-2025-people} is primarily derived from the article domain. However, for likely to be LLM-generated comments, the proportion of proper punctuation is high.

\begin{table}[!ht]
  \centering
  \footnotesize
  \caption{The proportion of LLM-generated code detected by DetectCodeGPT that forms code clones with code in GPTCloneBench. File-level clones refer to code clones found by comparing entire files containing code detected as likely to be LLM-generated with GPTCloneBench, while method-level clones refer to code clones obtained by splitting those files into individual methods for clone detection. Company-maintained repositories are shown above, and community-maintained repositories are shown below.
 }
  \label{tab:gpt-clone}
    \begin{tabular}{ccc}
    \toprule
    \textbf{Repos}
      & \textbf{File-level}
      & \textbf{Method-level}\\

    \midrule
    \textbf{Guava} & 0.54\%  & 0.05\%\\
    \textbf{Zap} & 0.00\% & 0.00\%\\

    \midrule
    \textbf{Jadx} & 0.57\% & 0.31\%\\
    \textbf{Kafka} & 3.09\% & 0.95\%\\
    \textbf{Pandas} & 0.00\% & 0.00\%\\
    
    \bottomrule
  \end{tabular}

\end{table}

\begin{table*}[!ht]
\centering
\begin{minipage}[]{0.48\textwidth}
  \centering
  \caption{Percentage of comments in company-maintained repositories detected as LLM-generated by different detectors across linguistic characteristics, using the average threshold.
  S represents Spelling
  ; P represents Punctuation
  ; G represents Grammar
  ; and V represents Vocabulary of AI-related word
  . Higher spelling, punctuation, and grammar percentage suggest more linguistically polished comments, while a higher vocabulary percentage indicates a higher Log-likelihood that the comment was generated by LLMs due to the frequent use of AI-related words. - indicates that no comments were detected as LLM-generated by a detector. 
  All numbers are in percentage.}
  \label{tab:criteria_company_avg}
  \footnotesize
  \begin{tabular}{cccccc}
    \toprule
    \textbf{Repos} & \textbf{Detectors} & \textbf{S} & \textbf{P} & \textbf{G} & \textbf{V}\\
    \midrule

    \multirow{8}{*}{\textbf{Go-github}} & Binoculars & 0.20 & 99.8 & 80.57 & 18.00\\
    & Detectgpt & 0.00 & 93.89 & 32.67 & 44.67\\
    & Entropy & 100 & 100 & 100 & 0.00\\
    & Fast-DetectGPT & 0.00 & 93.25 & 30.97 & 50.41\\
    & Log-likelihood & 0.05 & 97.74 & 55.23 & 20.40\\
    & Log-Rank & 0.32 & 96.99 & 49.93 & 23.09\\
    & LRR & 0.06 & 96.53 & 44.26 & 18.69\\
    & Rank & 0.11 & 98.84 & 79.66 & 22.23\\
    \midrule
    \multirow{8}{*}{\textbf{Guava}} & Binoculars & 58.38  & 100 & 74.60 & 10.46\\
    & Detectgpt & 11.59 & 74.6 & 28.24 & 59.56\\
    & Entropy & 49.65 & 96.10 & 88.79 & 12.33\\
    & Fast-DetectGPT & 0.00 & 78.38 & 70.27 & 72.97\\
    & Log-likelihood & 32.97 & 92.19 & 44.38 & 18.12\\
    & Log-Rank & 29.08  & 91.07 & 39.36 & 35.33\\
    & LRR & 43.63 & 88.24 & 40.69 & 22.06\\
    & Rank & 55.05 & 90.38 & 53.12 & 28.12\\
    \midrule
    \multirow{8}{*}{\textbf{Liquid}} & Binoculars & 27.17 & 100 & 20.65 & 11.96\\
    & Detectgpt &  11.84 & 89.47 & 6.58 & 34.21\\
    & Entropy & - & - & - & -\\
    & Fast-DetectGPT & 0.00 & 100 & 0.00 & 8.7\\
    & Log-likelihood & 76.94 & 100 & 75.65 & 1.94\\
    & Log-Rank & 69.45 & 97.87 & 67.50 & 6.57\\
    & LRR & 0.00 & 100 & 0.00 & 8.00\\
    & Rank & 84.51 & 98.86 & 84.97 & 2.51\\
    \midrule
    \multirow{8}{*}{\textbf{Zap}} & Binoculars & 0.00 & 100 & 33.33 & 66.67\\
    & Detectgpt &  0.00 & 94.89 & 5.24 & 28.23\\
    & Entropy & 37.5 & 100 & 50.00 & 0.00\\
    & Fast-DetectGPT & 0.00 & 85.78 & 11.85 & 68.25\\
    & Log-likelihood & 0.00 & 94.84 & 4.60 & 26.36\\
    & Log-Rank & 0.00 & 94.19 & 6.34 & 28.27\\
    & LRR & 0.00 & 96.23 & 3.37 & 24.6\\
    & Rank & 0.00 & 96.81 & 3.42 & 23.46\\
    \midrule
    \multicolumn{2}{c}{\textbf{Average}} & 22.21 & 94.74 & 41.49 & 25.68\\

    \bottomrule
  \end{tabular}
\end{minipage}
\hfill
\begin{minipage}[]{0.48\textwidth}

  \centering
  \caption{Percentage of comments in community-maintained repositories detected as LLM-generated by different detectors across linguistic characteristics, using the average threshold.
  S represents Spelling
  ; P represents Punctuation
  ; G represents Grammar
  ; and V represents Vocabulary of AI-related word
  . Higher spelling, punctuation, and grammar percentage suggest more linguistically polished comments, while a higher vocabulary percentage indicates a higher Log-likelihood that the comment was generated by LLMs due to the frequent use of AI-related words. 
  - indicates that no comments were detected as LLM-generated by a detector. 
  All numbers are in percentage.}
  \label{tab:criteria_community_avg}
  \footnotesize
  \begin{tabular}{cccccc}
    \toprule
    \textbf{Repos} & \textbf{Detectors} & \textbf{S} & \textbf{P} & \textbf{G} & \textbf{V}\\
    \midrule

    \multirow{8}{*}{\textbf{Act}} & Binoculars & 74.16 & 100 & 77.00 & 1.81\\
    & Detectgpt & 0.00 & 84.94 & 3.77 & 69.04\\
    & Entropy & 52.94 & 95.29 & 81.18 & 8.82\\
    & Fast-DetectGPT &  25.53 & 97.87 & 27.66 & 31.91\\
    & Log-likelihood & 54.60 & 100 & 50.19 & 8.81\\
    & Log-Rank & 44.73 & 98.00 & 43.93 & 18.02\\
    & LRR & 63.48 & 98.09 & 60.08 & 10.83\\
    & Rank & 67.37 & 100 & 75.86 & 4.77\\
    \midrule
    \multirow{8}{*}{\textbf{Jadx}} & Binoculars & 93.03 & 100 & 92.13 & 0.22\\
    & Detectgpt & 81.82 & 97.73 & 82.95 & 9.09\\
    & Entropy & 61.33 & 99.84 & 83.28 & 6.72\\
    & Fast-DetectGPT & - & - & - & -\\
    & Log-likelihood & 62.50 & 100 & 65.62 & 9.38\\
    & Log-Rank & 92.13 & 98.69 & 92.65 & 2.62\\
    & LRR & 99.10 & 100 & 97.01 & 0.30\\
    & Rank & 96.98 & 100 & 96.70 & 0.55\\
    \midrule    
    \multirow{8}{*}{\textbf{Kafka}} & Binoculars & 60.69 & 93.86 & 78.54 & 19.19\\
    & Detectgpt & 9.41 & 79.24 & 21.66 & 50.02\\
    & Entropy & 72.07 & 99.29 & 91.97 & 9.60\\
    & Fast-DetectGPT & 34.15 & 93.90 & 70.73 & 36.59\\
    & Log-likelihood & 20.09 & 92.11 & 33.41 & 26.64\\
    & Log-Rank & 20.43 & 84.52 & 44.04 & 44.00\\
    & LRR & 41.58 & 84.65 & 56.93 & 30.20\\
    & Rank & 24.98 & 88.33 & 38.99 & 33.46\\
    \midrule
    \multirow{8}{*}{\textbf{Pandas}} & Binoculars & 41.37 & 93.88 & 62.59 & 16.91\\
    & Detectgpt & 0.38 & 79.72 & 3.50 & 70.59\\
    & Entropy & 52.89 & 99.41 & 90.34 & 8.01\\
    & Fast-DetectGPT & 0.00 & 75.68 & 1.24 & 82.63\\
    & Log-likelihood & 27.41 & 90.34 & 24.61 & 29.60\\
    & Log-Rank & 7.60 & 87.74 & 16.75 & 49.85\\
    & LRR & 55.56 & 100 & 38.89 & 3.70\\
    & Rank & 36.41 & 90.76 & 26.63 & 27.17\\
    \midrule
\multicolumn{2}{c}{\textbf{Average}} & 47.57 & 93.67 & 55.83 & 23.26\\

    \bottomrule
  \end{tabular}
  \end{minipage}
\end{table*}

\begin{tcolorbox}[colback=black!5!white, colframe=black, title=Answer to RQ2, fonttitle=\bfseries]
The code detected as likely to be LLM-generated contains code clones within their own repositories with majority of the repositories exceeding 70\% at file-level and 20\% at method-level, whereas almost few code is detected as likely to be LLM-generated has clones in GPTCloneBench. Unlike the previous work reporting on the properties of text generated by LLMs, the comments detected as likely to be generated by LLMs show a high proportion of correct punctuation usage, but a low frequency of AI-related vocabulary and correct grammar.

\end{tcolorbox}

\subsection{RQ3: How does LLMs usage detected by the detectors differ between community-maintained and company-maintained repositories? }

The company-maintained repositories showed a higher average proportion of likely to be LLM-generated content across code and comment, as detected by the detectors. As shown in Figure \ref{fig:detectcodegpt_result}, the company-maintained repositories generally exhibited a higher proportion of code detected as likely to be LLM-generated. However, the Liquid repository showed very low percentages of code detected as likely to be LLM-generated. Among the community-maintained repositories, the proportion of code detected as likely to be LLM-generated was around 20\%, which showed a lower percentage.

\yj{As illustrated in Figure \ref{fig:comment_avg}, although the different detectors showed different results, the highest proportions of comments detected as likely to be LLM-generated in company-maintained repositories were higher than those in community-maintained repositories. For example, Log-Rank, LRR and Log-likelihood showed a high proportion of comment detected as likely to be LLM-generated, with values exceeding 80\% for the Go-github. For the Zap repository, Log-Rank, Log-likelihood and DetectGPT detected almost all comments (around 100\%) as likely to be LLM-generated. In the Liquid repository, some detectors identify more than 50\% of the comments as likely to be LLM-generated.
In community-maintained repositories, nearly all the repositories had less than 70\% of comments detected as likely to be LLM-generated by all detectors. For the Kafka, and Pandas repositories, all detectors detected less than 70\% of the comments as likely to be LLM-generated. In the Jadx repository, the proportion of comments detected as likely to be LLM-generated was below 40\% for all detectors except Entropy. Similarly, In the Act repository, the proportion of comments detected as likely LLM-generated remains below 60\% for all detectors, except for Entropy in 2021 and 2022. }

As shown in Table \ref{tab:code-clone}, company-maintained repositories have a higher percentage of likely to be LLM-generated code containing file-level code clones, with the Go-github, Guava, Liquid, and Zap repositories all exceeding 90\%. In contrast, the Kafka and Pandas repositories, which are community-maintained repositories, showed relatively lower percentages of likely to be LLM-generated code containing file-level code clones. In particular, the Pandas repository has only 45.32\% in the intra-repository file-level code clone comparison. In terms of intra-repository method-level comparisons, the proportions are quite similar between company-maintained and community-maintained repositories. Company-maintained repositories show relatively higher code clone percentages, with the lowest at 26.3\% and the highest at 34.98\%, while community-maintained repositories have relatively lower percentages, with the lowest at 14.55\% and the highest at 31.89\%. When compared with GPTCloneBench, both company-maintained and community-maintained repositories had a very low percentage of code detected as likely to be likely LLM-generated that contained code clones. For the characteristics of comments detected as likely to be LLM-generated, as shown in Tables \ref{tab:criteria_company_avg}, and \ref{tab:criteria_community_avg}, community-maintained repositories tend to be more linguistically correct and cleaner compared with company-maintained repositories.

As reported in Table \ref{tab:coding-process}, although both company-maintained and community-maintained repositories tended to have relatively high proportions of code detected as likely to be LLM-generated in the Tests category, the overall distribution across categories differed between company- and community-maintained repositories. In the company-maintained repositories, the Core logic category has a higher proportion of code detected as likely to be LLM-generated, while in the community-maintained repositories, it showed a more distributed pattern across the categories. For comments detected as likely to be LLM-generated, the company-maintained repositories showed a higher concentration in the Explanation or Meta categories. For example, the Go-github and Guava repositories showed 63\% and 67\% of comments in the Explanation category, respectively, and the Liquid and Zap repositories showed 74\% and 71.5\% in the Meta category respectively. In comparison, the community-maintained repositories showed a more varied pattern. For example, in the Pandas repository, 43.5\% of the comments detected as likely to be LLM-generated belonged to the PastFuture category.

\begin{tcolorbox}[colback=black!5!white, colframe=black, title=Answer to RQ3, fonttitle=\bfseries]
Compared with community-maintained projects, 
company-maintained repositories showed a higher proportion of code and comments detected as likely to be LLM-generated and
a higher proportion of intra-repository code clones in the code detected as likely to be LLM-generated.
Majority of the code detected as likely to be LLM-generated in the company-maintained repositories are test cases.
\end{tcolorbox}

\subsection{RQ4: \yj{How likely are bugs associated with LLM-generated code?} }


We found that only a small portion of the PreciseBugs dataset was detected as likely to be LLM-generated. Specifically, in the NVD data source, 10.79\% of the human-labelled bugs are detected as likely to be LLM-generated. Among these, the highest percentage of the human-labelled bugs (12.57\%) are in the CWE-287 type (e.g., a system failing to verify a user’s claimed identity), followed by 10.34\% in the CWE-129 type (e.g., a program using an untrusted array index without proper validation). In the OSS-Fuzz data source, 5.56\% of the human-labelled bugs are detected as likely to be LLM-generated, with 100\% in the Uncaught exception type (e.g., an exception that is thrown but not handled).

\begin{tcolorbox}[colback=black!5!white, colframe=black, title=Answer to RQ4, fonttitle=\bfseries]

Only a small percentage of the human-labelled bugs (10.79\% and 5.56\% from NVD and OSS-Fuzz, respectively) in repositories analyzed by DetectCodeGPT are likely to be generated by LLMs.

\end{tcolorbox}

\section{Discussion}
\label{section:discussion}

\noindent\yjnew{\textbf{Trends of LLM-Generated Content}}

\yjnew{Our results should be interpreted as proxy-based observations derived from detector outputs, rather than direct measurements of LLM usage. Given the absence of ground truth labels in repository data, the reported proportions reflect detector behaviour under specific thresholds and configurations.}
\yjnew{Moreover, we emphasize that high detection rates in early years should not be interpreted as evidence of actual LLMs adoption in software development workflows. The high detection rates may be driven by several factors, including detector bias and false positives, repetitive or template-based code patterns, retrospective repository modifications and structural similarities between human-written test code and LLM-generated outputs. Therefore, we treat the early year detection rates as a baseline for comparison, rather than as meaningful indicators of absolute LLM usage. Our analysis focuses on relative changes over time, rather than the absolute values. }
\yjnew{The observed decreasing trend in the proportion of code detected as likely to be LLM-generated may reflect multiple factors rather than a single underlying cause. One possible interpretation is that it may also reflect changes in developer usage patterns.} Developers may begin to use LLMs more strategically. Instead of relying on LLMs to generate code without modification, developers may employ it for tasks such as code completion, which is not the focus in this work. \yjnew{Another possible explanation is that newer generations of LLM-generated code may increasingly resemble human-written code, reducing the sensitivity of existing detectors over time.}

As LLMs rapidly improve and become more widely adopted,
our observation may not change significantly. As found in the previous work, code generated by LLMs contain bugs \cite{liu2024refining, tambon2025bugs, dakhel2023github} and requires developers to review the code generated by LLMs. In our study, we observed that the proportion of code detected as likely to be LLM-generated decreased over time in 50\% of the repositories (this is despite more advanced LLMs were introduced in the latter years), while the proportion of comments detected as likely to be LLM-generated have negligible changes over time. We believed this pattern is likely to persist because, even with the introduction of more advanced LLMs, developers still need to carefully review the code generated by LLMs. Comments detected as likely to be LLM-generated are mainly found in the Meta or Explanation categories. The use of LLMs to generate comments remained stable, and we expect this trend to continue in the future.

\begin{figure*}[!t]
    \centering
    \subfigure[Threshold at -20\% of the baseline]{
        \begin{minipage}[b]{0.3\textwidth}
            \centering
            \includegraphics[width=\textwidth]{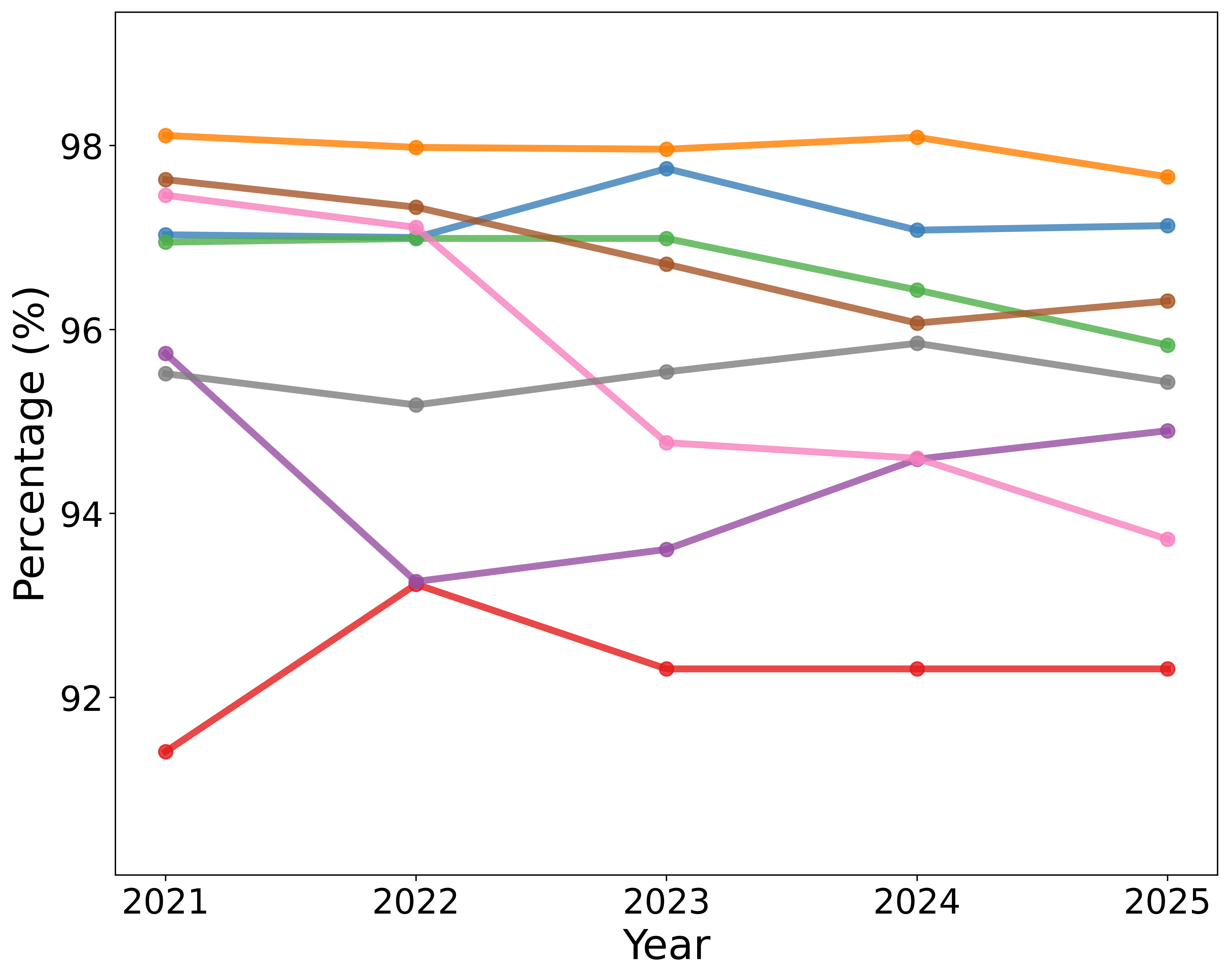}
            \vspace{-1em}
            \label{fig:-20_threshold}
        \end{minipage}
    }
    \hfill
    \subfigure[Baseline threshold]{
        \begin{minipage}[b]{0.3\textwidth}
            \centering
            \includegraphics[width=\textwidth]{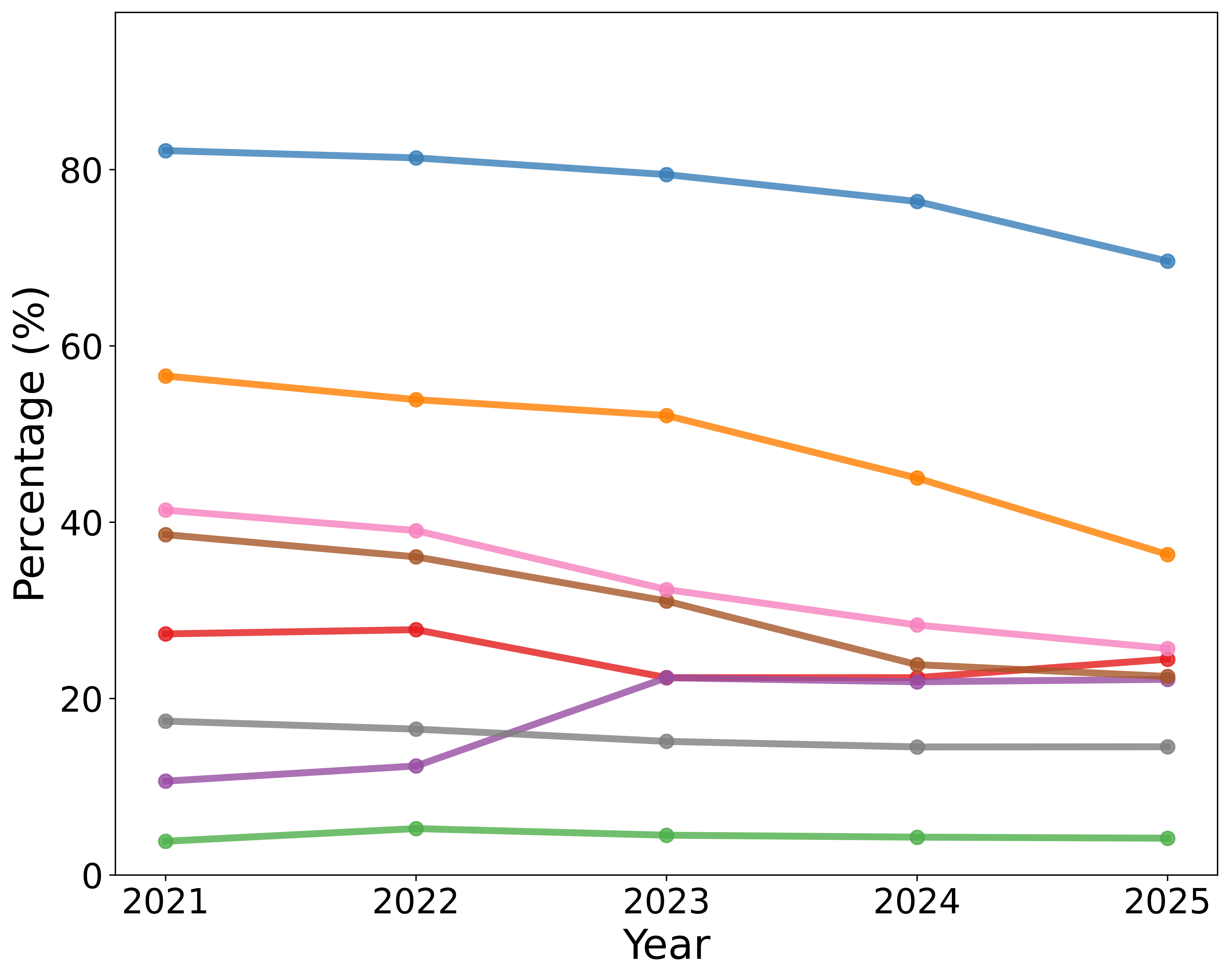}
            \vspace{-1em}
            \label{fig:base_threshold}
        \end{minipage}
    }
    \hfill
    \subfigure[Threshold at +20\% of the baseline]{
        \begin{minipage}[b]{0.3\textwidth}
            \centering
            \includegraphics[width=\textwidth]{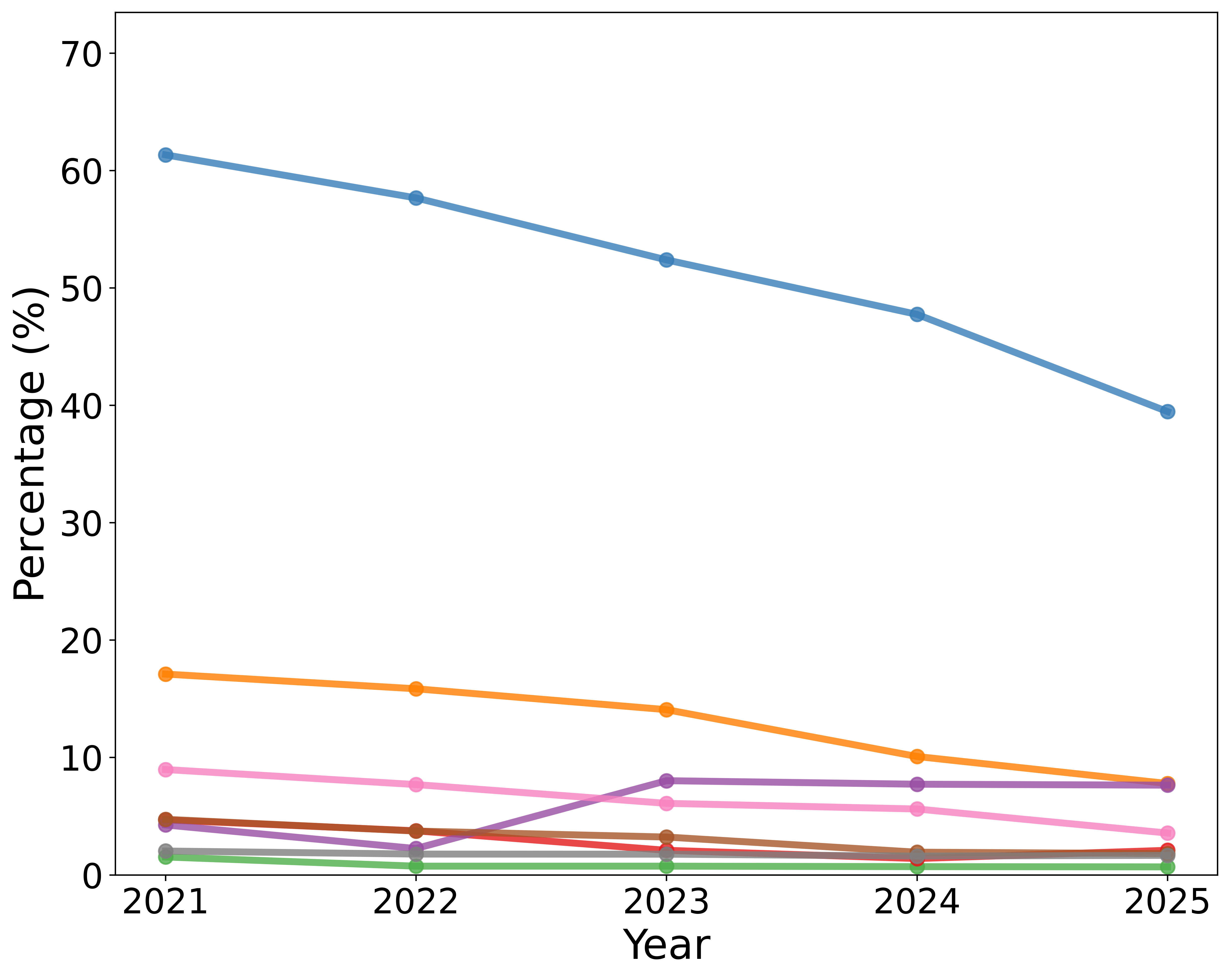}
            \vspace{-1em}
            \label{fig:20_threshold}
        \end{minipage}
    }

    \begin{minipage}{0.7\textwidth}
        \centering
        \includegraphics[width=\textwidth]{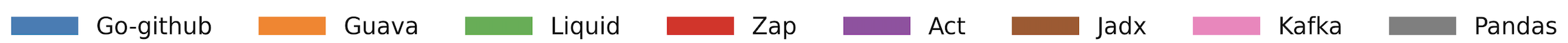}
    \end{minipage}

    \caption{Proportion of code detected as LLM-generated across different repositories, comparing the selected baseline threshold with variations of $-20\%$ and $+20\%$. Figure \ref{fig:base_threshold} shows the overall change in the proportion of code detected as likely LLM-generated using the baseline threshold, with +20\% and -20\% variations illustrated in Figures \ref{fig:20_threshold} and \ref{fig:-20_threshold}, respectively.}

    \label{fig:code_threshold}
\end{figure*}

\begin{figure*}[!t]
    \centering
    \subfigure[Threshold at -20\% of the baseline]{
        \begin{minipage}[b]{0.3\textwidth}
            \centering
            \includegraphics[width=\textwidth]{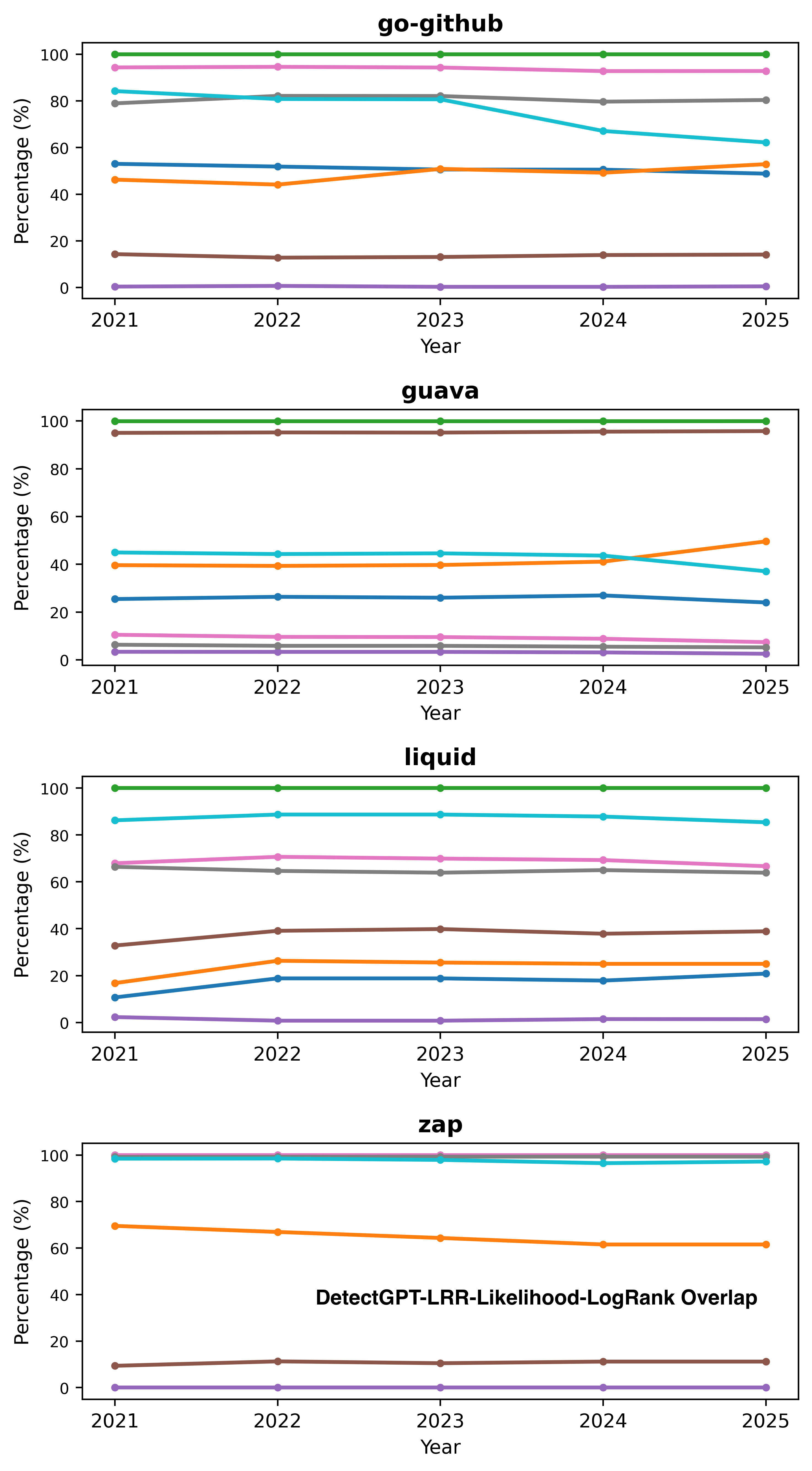}
            \vspace{-1em}
            \label{fig:comment_minus_company}
        \end{minipage}
    }
    \subfigure[Baseline threshold]{
        \begin{minipage}[b]{0.3\textwidth}
            \centering
            \includegraphics[width=\textwidth]{plot/comments_group_1.png}
            \vspace{-1em}
            \label{fig:comment_se_company}
        \end{minipage}
    }%
    \subfigure[Threshold at +20\% of the baseline]{
        \begin{minipage}[b]{0.3\textwidth}
            \centering
            \includegraphics[width=\textwidth]{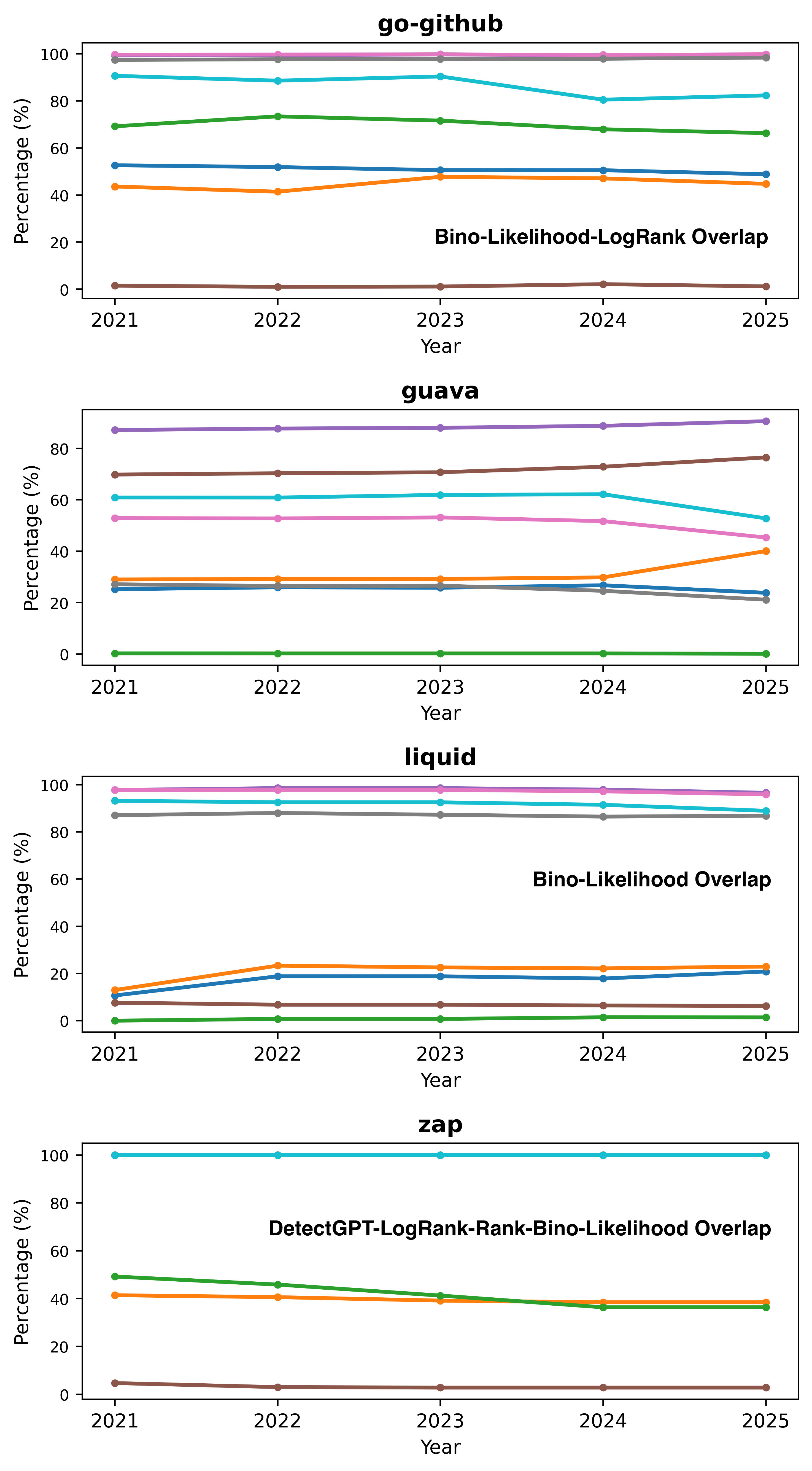}
            \vspace{-1em}
            \label{fig:comment_plus_company}
        \end{minipage}
    }

    \begin{minipage}{0.7\textwidth}
        \centering
        \includegraphics[width=\textwidth]{plot/legend.png}
    \end{minipage}

    \caption{\yjnew{Proportion of comment  detected as LLM-generated in company-maintained repositories, comparing the AISE threshold with variations of $-20\%$ and $+20\%$. Figure \ref{fig:comment_se_company} shows the overall change in the proportion of comment detected as likely LLM-generated using the AISE dataset threshold, with +20\% and -20\% variations illustrated in Figures \ref{fig:comment_plus_company} and \ref{fig:comment_minus_company}, respectively. The overlapped lines are mentioned in wordings in the plots.
    The following detector lines overlap in the corresponding repositories: in Figure \ref{fig:comment_minus_company}: in Zap: DetectGPT, LRR, Log-Likelihood and Log-Rank. in Figure \ref{fig:comment_plus_company}: in Go-github: Binoculars, Log-Likelihood and LogRank; in Liquid: Binoculars and Log-Likelihood; in Zap: DetectGPT, LogRank, Rank, Binoculars and Log-Likelihood. }}

    \label{fig:comment_theshold_company}
\end{figure*}

\begin{figure*}[!t]
    \centering
    \subfigure[Threshold at -20\% of the baseline]{
        \begin{minipage}[b]{0.3\textwidth}
            \centering
            \includegraphics[width=\textwidth]{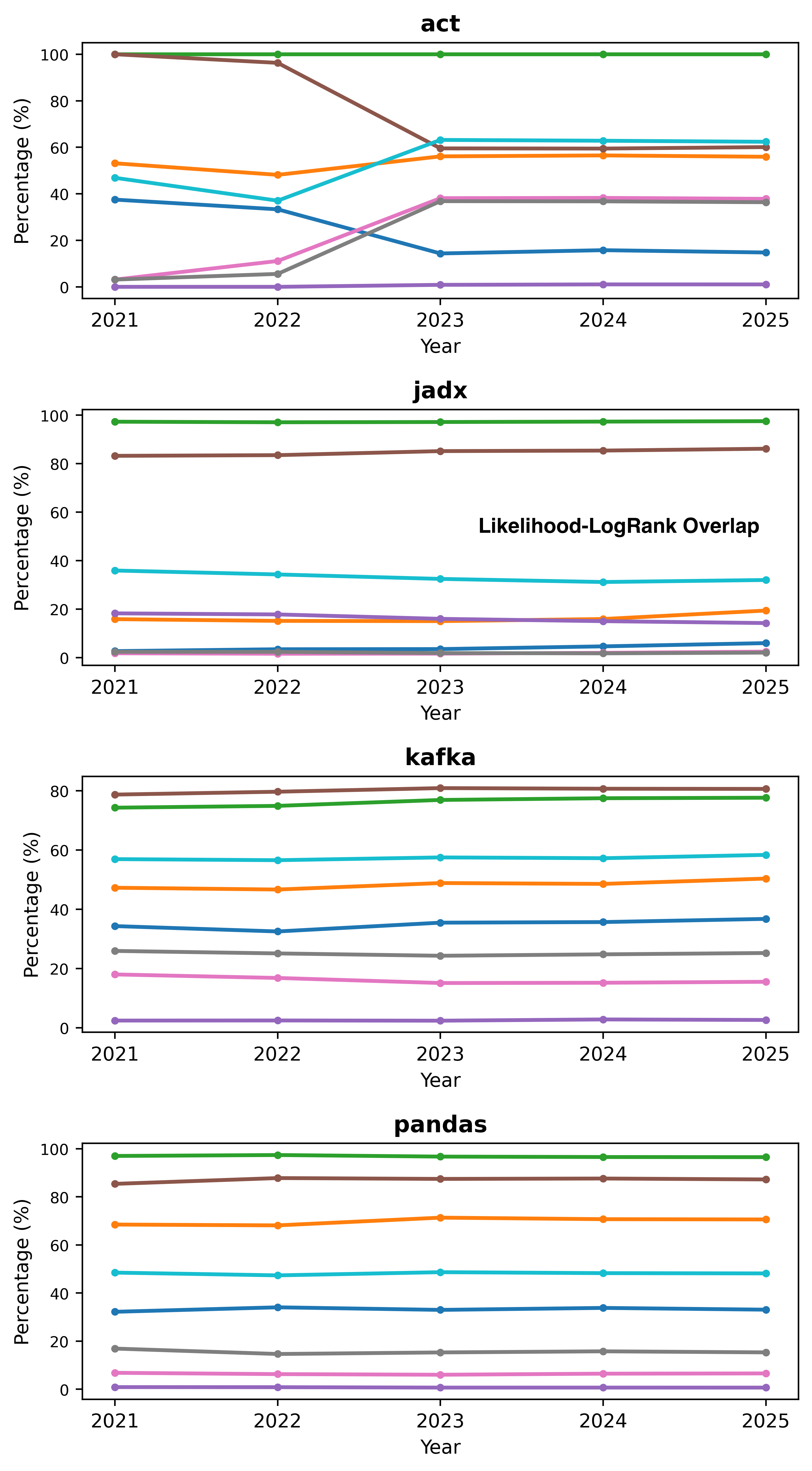}
            \vspace{-1em}
            \label{fig:comment_minus_community}
        \end{minipage}
    }
    \subfigure[Baseline threshold]{
        \begin{minipage}[b]{0.3\textwidth}
            \centering
            \includegraphics[width=\textwidth]{plot/comments_group_2.png}
            \vspace{-1em}
            \label{fig:comment_se_community}
        \end{minipage}
    }%
    \subfigure[Threshold at +20\% of the baseline]{
        \begin{minipage}[b]{0.3\textwidth}
            \centering
            \includegraphics[width=\textwidth]{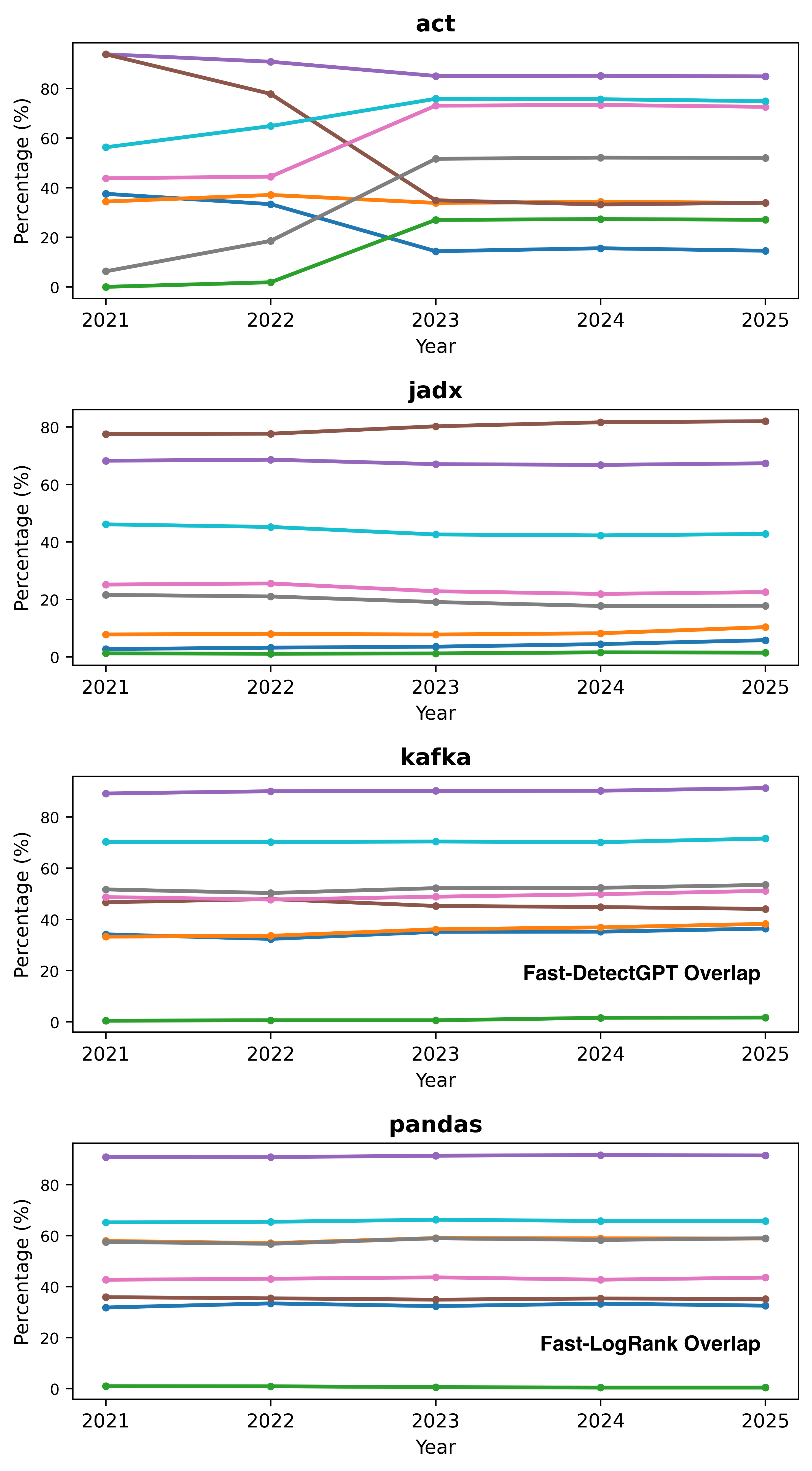}
            \vspace{-1em}
            \label{fig:comment_plus_community}
        \end{minipage}
    }

    \begin{minipage}{0.7\textwidth}
        \centering
        \includegraphics[width=\textwidth]{plot/legend.png}
    \end{minipage}

    \caption{\yjnew{Proportion of comment detected as LLM-generated in community-maintained repositories, comparing the AISE threshold with variations of $-20\%$ and $+20\%$. Figure \ref{fig:comment_se_community} shows the overall change in the proportion of comment detected as likely LLM-generated using the AISE dataset threshold, with +20\% and -20\% variations illustrated in Figures \ref{fig:comment_plus_community} and \ref{fig:comment_minus_community}, respectively. The overlapped lines are mentioned in wordings in the plots.
    The following detector lines overlap in the corresponding repositories: in Figure \ref{fig:comment_minus_community}: in Jadx: Log-Likelihood and Log-Rank. In Figure \ref{fig:comment_plus_community}: in Kafka: Fast-DetectGPT and DetectGPT; in Pandas: Fast-DetectGPT and LogRank.}}

    \label{fig:comment_theshold_community}
\end{figure*}

\hfil\break
\noindent\yjnew{\textbf{Detector Limitations and Measurement Bias}}

\yjnew{Our detector-based approach may underestimate the actual usage of LLMs in software repositories. }
\yj{In earlier years, LLM-generated code may have been easier to detect, while in recent years, it may come from more advanced models, which could lead us to underestimate LLMs usage.} In addition, in many cases, the LLM-generated code may undergo substantial human revision before committing to the repositories, making it difficult for detectors to identify it as LLM-generated. 

Meanwhile, detectors may also overestimate LLM-generated code and comments. \yjnew{Human-written code with repetitive or highly structured patterns may resemble generated code and may be incorrectly detected as likely LLM-generated. Since we did not conduct surveys or interviews with developers, we could not validate whether detected content was generated with LLMs.} We chose not to conduct such a survey to investigate how developers are using LLMs to generate code and comments in their repositories, because we believed that these inquiries could raise ethical concerns such as identifying a specific developer \cite{tahaei2024surveys}. \yjnew{Consequently, our findings should be interpreted as detector-based proxy observations rather than precise measurements of actual LLM usage.}

\yj{To evaluate the robustness of these observations, we conducted a sensitivity analysis by varying the threshold DetectCodeGPT by $\pm 20\%$} \yjnew{ and the thresholds of the comment detectors by $\pm 20\%$. The results for code are shown in Figure \ref{fig:code_threshold}, while the results for comments are shown in Figures \ref{fig:comment_theshold_company} and \ref{fig:comment_theshold_community}. For code detection, adjusting} the DetectCodeGPT threshold from 1.3 by $\pm 20\%$ causes notable fluctuations in the proportions. However, the overall trend remains similar. For most repositories, the proportion of code detected with the baseline threshold and the +20\% variation decreases over time. With a -20\% variation, the threshold is generous that over 90\% of code is detected as likely LLM-generated across all repositories. \yjnew{For comment detection, different detectors produced different absolute proportions under threshold variations. Nevertheless, the overall patterns remained relatively stable across repositories. These findings suggest that our observations are relatively robust. However, even thresholds adapted using the AISE dataset should still be regarded as approximations rather than definitive decision thresholds. In addition, constructing ground-truth datasets for developer comments remains an open challenge, as it is difficult to obtain verified human-written and LLM-generated comments.}

\hfil\break
\noindent\yjnew{\textbf{Characteristics of Likely to be LLM-Generated Content}}

\yj{We observed that the majority of code detected as likely to be LLM-generated is concentrated in the Tests category. Through the coding process, we observed that code, detected as likely to be LLM-generated in Tests category, consist of highly repetitive code, hard-coded JSON strings, and symmetric benchmark loops. Because DetectCodeGPT perturbs input code by strategically inserting spaces and newlines \cite{10.1109/ICSE55347.2025.00005}, these perturbations may not significantly shift the perplexity, resulting in low perturbation sensitivity for such code. As a result, even human-written test code can be detected as likely to be LLM-generated because its structural patterns are likely to be detected as LLM-generated.}

\yj{In addition, we found that likely LLM-generated code exhibited a high intra-repository clone rate. This indicates a risk of technical debt. We recommend that automated refactoring tools can be integrated into the CI/CD pipeline to detect and consolidate repetitive LLM-generated snippets, preventing long-term maintenance issues.}

We found that the company-maintained repositories contained a higher proportion of LLM-generated code and comments compared to community-maintained repositories. This may be because companies tend to employ LLMs to accelerate project development and maintain internal consistency, which in turn leads to higher percentage of code and comments detected as likely to be LLM-generated.

Previous studies have found that developers are often reluctant to use LLMs because the generated code does not meet their requirements, the output is difficult to control, and they spend a lot of time debugging. However, our results showed that only a small percentage of the human-labelled bugs (10.79\% and 5.56\% from NVD and OSS-Fuzz, respectively) in repositories were likely to be generated by LLMs. \yj{While DetectCodeGPT may underestimate LLM-generated code that contains bugs, the low percentage suggests that developers may have modified the LLM-generated code to remove any bugs before committing to the repositories.} \yj{It should be noted that the bugs in our analysis, sourced from NVD and OSS-Fuzz, primarily represent security vulnerabilities and runtime exceptions. Therefore, the results may not generalize to all types of software defects.}

\section{Threats to Validity}
\label{section:threats}

In this section, we address threats to the validity of our study.

\subsection{Internal Validity}
In our experiments, we used existing detectors to identify LLM-generated content in repositories. However, the repositories do not have ground truth on whether the content is LLM-generated. Therefore, our analysis relies on the existing detectors' outputs, which may not perfectly reflect the true extent of developers' LLMs usage, as they may overestimate or underestimate the LLM-generated code and comments. To mitigate this, we made every effort to use the existing resources to detect the code and comments that were likely to be generated by LLMs by using multiple detectors and threshold settings, improving the reliability of our measurements. 
\yjnew{ Moreover, we conducted a sensitivity analysis with $\pm 20\%$ threshold variation. The results show that while absolute values change, the overall temporal trends remain stable.}



\subsection{External Validity} 
In this study, we primarily focused on repositories maintained by large companies and active communities, which have significant contributions and a high number of stars. \yj{Although these projects are active, the results are skewed toward mature and popular ecosystems, and thus may not generalize to other ecosystems.}  Moreover, we selected repositories developed in programming languages that can be detected by DetectCodeGPT. As a result, the generalizability of our findings may be limited to similar types of repositories, and may not directly extend to less active repositories, smaller projects, or repositories in unsupported programming languages. \yj{Finally, our dataset is restricted to open-source projects. Therefore, the results cannot be generalized to closed-source industrial repositories.} However, to make our findings more representative, we analyzed repositories from different companies and organizations. These repositories were also developed in different programming language to improve the diversity of our dataset.

\subsection{Construct Validity.}

Although we observed the percentage of the code and human-labelled bugs detected as likely to be LLM-generated is low, developers may still use LLMs to assist them on code refactoring, debugging, code comprehension, and other tasks, and that the developers may also rewrite or paraphrase the content generated by LLMs and commit into the repositories. In such cases, we believe there is no one single detectors that can identify such LLM-assisted content that has been substantially modified. Furthermore, there is no existing study that reports how much of a modified LLM-generated content can be identified by these detectors. To mitigate this, we use a variety of detectors of various abilities, including state-of-the-art detectors, to detect content as likely to be LLM-generated. \yjnew{In addition, the detection of LLM-generated comments introduces further construct validity challenges due to domain differences. Programming comments often contain inline code, parameter descriptions, abbreviations, and DSL-like structures that differ substantially from the natural language data commonly used to calibrate existing detectors. As a result, natural language detectors may not generalize perfectly to software engineering comments and may introduce domain-specific detection bias.}

\yjnew{\subsection{Detection Validity.}}

\yjnew{Our analysis relies on zero-shot detectors whose thresholds are derived from proxy datasets rather than ground-truth annotations. As optimal thresholds vary across domains, the classification of LLM-generated content remains uncertain. To mitigate this, we applied domain-specific threshold derivation and sensitivity analysis. However, the results should still be interpreted as indicative rather than definitive. In addition, detector behaviour may change over time as newer LLMs evolve. Detectors calibrated on earlier generations of LLM outputs may become less effective when applied to newer models whose outputs increasingly resemble human-written content.}

\section{Conclusion}
\label{section:conclusion}

With the advancement of LLMs, their adoption has become increasingly popular among developers. In this study, we aim to \yjnew{show} how \yjnew{code and comments detected as likely to be LLM-generated appear in repositories}. We analyzed active repositories maintained by both companies and the community. Our \yjnew{proxy-based observations indicate}  that the proportion of likely to be LLM-generated code decreased over time, while the proportion of comments detected as likely to be LLM-generated remained stable. \yjnew{Based on detector-based proxy analysis,} we also \yjnew{observed} that company-maintained repositories had a higher percentage of code and comments detected as likely to be LLM-generated and exhibited higher intra-repository code clone rates, whereas community-maintained projects produced linguistically cleaner comments detected as likely to be LLM-generated. In addition, only a small percentage of the human-labelled bugs (10.79\% and 5.56\% from NVD and OSS-Fuzz, respectively) in the repositories were detected as likely to be LLM-generated. 
Building on this study, future work could \yj{establish guidelines for reviewing, such as designating low-priority test files for AI-assisted review, to improve efficiency.}


\section*{Acknowledgement}
Calculations were performed using the Sulis Tier 2 HPC platform hosted by the Scientific Computing Research Technology Platform at the University of Warwick. Sulis is funded by EPSRC Grant EP/T022108/1 and the HPC Midlands+ consortium.

\section*{Declaration of generative AI and AI-assisted technologies in the manuscript preparation process}

During the preparation of this work the author(s) used detectors (including Binoculars, Log-Likelihood, Entropy, Rank, Log-Rank, LRR, DetectGPT, Fast-DetectGPT and DetectCodeGPT) in order to run the core experiments required in the manuscript. We did not use generative AI or AI-assisted technologies to produce any of the content in the manuscript. 


\bibliographystyle{IEEEtran}
\bibliography{bibliography} 

\end{document}